\documentclass[%
aip,jcp,amsmath,amssymb, reprint, twocolumn
]{revtex4-2}
\usepackage{amsmath}
\usepackage{mathtools}
\usepackage{float}
\usepackage{amsfonts}
\usepackage{amssymb}
\usepackage{dcolumn} 
\usepackage{array}
\newcolumntype{P}[1]{>{\centering\arraybackslash}p{#1}}
\newcolumntype{M}[1]{>{\centering\arraybackslash}m{#1}}
\newcolumntype{C}[1]{>{\centering\arraybackslwash}p{#1}}
\usepackage{float}
\usepackage{psfrag}
\usepackage{tabularx}
\usepackage{stackengine}
\usepackage{amssymb}
\usepackage{mathtools}  
\usepackage{xfrac} 
\usepackage[T1]{fontenc}
\usepackage{graphicx}
\usepackage{diagbox}
\usepackage[caption=false]{subfig}
\usepackage{tikz}
\usetikzlibrary{positioning}
\usetikzlibrary{arrows}
\usetikzlibrary{trees}
\usepackage{amssymb}
\usetikzlibrary{decorations.pathmorphing}
\usetikzlibrary{decorations.markings}
\usetikzlibrary{automata,positioning}
\usepackage{braket}
\usepackage{rotating}
\usepackage{adjustbox}
\usepackage{ragged2e}
\usepackage{simplewick}
\usepackage{simpler-wick}

\usepackage{csquotes}
\usepackage{physics,amsmath}
\usepackage{xcolor}
\usepackage{fancyhdr}
\UseRawInputEncoding
\usepackage[normalem]{ulem}

\usepackage{scalerel}
\usepackage{hyperref}
\usepackage{appendix}
\usepackage{multirow}
\usepackage[colorinlistoftodos]{todonotes}

\begin{document}

\author{Chayan Patra}
\affiliation{ Department of Chemistry,  \\ Indian Institute of Technology Bombay, \\ Powai, Mumbai 400076, India}
\author{Dibyendu Mondal}
\affiliation{ Department of Chemistry,  \\ Indian Institute of Technology Bombay, \\ Powai, Mumbai 400076, India}
\author{Sonaldeep Halder}
\affiliation{ Department of Chemistry,  \\ Indian Institute of Technology Bombay, \\ Powai, Mumbai 400076, India}
\author{Dipanjali Halder}
\affiliation{Department of Chemistry,\\ Technical University of Denmark,\\ Kemitorvet Building 207, DK-2800
Kongens Lyngby, Denmark.}
\author{Mostafizur Rahaman Laskar}
\affiliation{Quantum Division, IBM Research Lab, Bangalore, India}
\author{Richa Goel}
\affiliation{Quantum Division, IBM Research Lab, Gurgaon, India}
\author{Rahul Maitra}
\email{rmaitra@chem.iitb.ac.in}
\affiliation{ Department of Chemistry,  \\ Indian Institute of Technology Bombay, \\ Powai, Mumbai 400076, India}
\affiliation{Centre of Excellence in Quantum Information, Computing, Science \& Technology, \\ Indian Institute of Technology Bombay, \\ Powai, Mumbai 400076, India}


\title{Physics-Informed Generative Machine Learning for Accelerated Quantum-Centric Supercomputing\\
}





\begin{abstract}
Quantum centric supercomputing (QCSC) framework, such as sample-based quantum diagonalization (SQD) holds immense promise toward achieving practical quantum utility to solve
classically hard sampling problems in quantum chemistry.
QCSC leverages quantum computers to perform the classically intractable task of sampling the dominant fermionic configurations from the Hilbert space that have substantial support to a target state, followed by Hamiltonian diagonalization on a classical processor. However, noisy quantum hardware produces erroneous samples upon measurements, making robust and efficient configuration-recovery strategies essential for a scalable QCSC pipeline. Toward this, in this work, we introduce PIGen-SQD, an efficiently designed QCSC workflow that utilizes the capability of generative machine learning (ML) along with physics-informed configuration screening via implicit low-rank tensor decompositions for accurate fermionic state reconstruction. The physics-informed pruning is based on a class of efficient perturbative measures that, in conjunction with samples drawn from quantum hardware,
provide a substantial overlap with the target state. This distribution induces an anchoring effect on the generative ML models to stochastically explore only the dominant sector of the Hilbert space for effective identification of additional important configurations in a self-consistent manner. Our numerical experiments performed on IBM Heron R2 and R3 quantum processors with up to 52 qubit experiments demonstrate this synergistic workflow produces compact, high-fidelity subspaces that substantially reduce diagonalization cost while maintaining chemical accuracy under strong electronic correlations. 
With such a concerted integration of classical many-body intuitions, generative ML and quantum computers as a sampling engine, PIGen-SQD
offers a promising pathway toward accurate and systematically improvable quantum simulations on utility-scale quantum  hardware.

\end{abstract}

\maketitle


\section{Introduction}

Quantum computing holds immense promise for efficiently solving complex and classically intractable problems.
A key application area is quantum chemistry, which seeks to determine the ground and excited state energies of many-body fermionic systems by solving the
time-independent Schr\"odinger equation within the Born-Oppenheimer approximation.
Designing efficient and accurate quantum algorithms is key toward solving some of the open problems in chemistry such as computation of
ligand binding affinity of a drug molecule at a given protein target site\cite{santagati2024drug,gunther2025use,baiardi2023quantum}, accurate electronic structure calculations of the Fe-S clusters
for a better understanding of biological nitrogen fixation under ambient conditions\cite{montgomery2018strong}, or designing high temperature superconducting materials, which involve strong many-body correlation effects driven by Cu 3d electron bands\cite{chan2024spiers}.
However, with the current degree of hardware noise, quantum algorithms like the quantum phase estimation
(QPE)\cite{abrams1997simulation,abrams1999quantum} is not viable due to its requirement of deep circuits and long coherence time. Consequently, near-term research focuses on hybrid quantum-classical
approaches like the variational quantum eigensolver (VQE)\cite{peruzzo2014variational, cerezo2021variational, bharti2022noisy}, quantum approximate optimization algorithm (QAOA)\cite{farhi2014quantum} and projective quantum eigensolver (PQE)\cite{stair2021simulating} where
quantum processors are dedicated to the evaluation of expectation values and classical processors perform parameter optimization. However, optimization of such hybrid
algorithms for practically relevant systems is not guaranteed due to Barren Plateaus\cite{mcclean2018barren, cerezo2021cost,larocca2025barren} and far-from-optimal local traps\cite{anschuetz2022quantum} in the optimization landscape, prolonged optimization
runtimes and steep measurement overhead.
Reducing the resource requirements for achieving practical quantum advantage of such hybrid algorithms is still an active area of research\cite{grimsley2019adaptive,smart2021quantum,Grimsley2023,ryabinkin2018qubit,bauman2019downfolding,matsuzawa2020jastrow,ryabinkin2020iterative,yordanov2020efficient,kowalski2021dimensionality,mondal2023development,halder2022dual,claudino2021improving,halder2023corrections,halder2023measurement,kowalski2023quantum,burton2023exact,burton2024accurate,halder2024noise,halder2023machine,patra2024projective,patra2024toward,patra2025energy, haidar2025non}.
Despite such intensive research efforts, scaling up electronic structure calculations solely with
such variational optimization algorithms remains a nontrivial challenge.
In this scenario,
alternative algorithmic developments are required to scale up quantum chemical calculations using quantum computers
for practical quantum advantage.

Toward this, recently a distinct class of hybrid architecture termed Quantum-Centric Supercomputing (QCSC) has garnered attention
due to its enhanced noise resilience and capability of sampling classically hard probability distributions, presenting a potential path forward for larger-scale simulations. QCSC is an emerging paradigm characterized by the seamless integration of quantum computing resources with traditional High-Performance Computing (HPC) infrastructures.
This unified architecture is designed to address highly complex, real-world computational problems by distributing the workload to optimally leverage the distinct advantages of both quantum and classical processors.
Some of the recently developed QCSC algorithms are inspired by classical selected configuration interaction (SCI)\cite{sherrill1999configuration, huron1973iterative,holmes2016heat,liu2016ici,sharma2017semistochastic} such as quantum SCI (QSCI)\cite{kanno2023quantum},
sample-based quantum diagonalization (SQD)\cite{robledo2025chemistry, yoshioka2025krylov}
and its different variants\cite{kanno2023quantum,robledo2025chemistry,barison2025quantum,yoshioka2025krylov,shajan2025toward,piccinelli2025quantum,danilov2025enhancing,yu2025quantum}.
In such QCSC methods, generally a state is prepared on a quantum computer that approximately represents the target state and repeated measurements are performed to obtain samples that produce a set of computational basis states.
For a molecular system these sampled computational basis states represent
important electronic configurations which have dominant contribution to the target electronic state.
HPC is required post-sampling for projection
of the Hamiltonian in this electronic configuration subspace and diagonalization to get the energy of the approximate state.
However, due to the presence of hardware noise, the sampled distributions often contain a substantial number of states that violate fundamental physical symmetries of the molecular system, such as particle number or spin
conservation, leading to erroneous results upon diagonalization.
To filter these noise-induced configurations and faithfully recover the target state, SQD employs a configuration recovery (CR) module.
CR expands the subspace to generate more dominant electronic configurations
in a self-consistent manner based on the average occupancy of the corresponding spin-orbitals with subsequent diagonalization.
Since sampling dominant configurations from Hilbert space is a computationally challenging task, SQD has the potential to
provide advantage over classical SCI methods, provided it produces accurate energy estimations using comparatively reduced
dimensional configuration subspace for diagonalization.
While SQD with its configuration recovery can produce accurate results for chemically complex systems
under strong electronic correlations\cite{robledo2025chemistry}, the accuracy and efficacy depend on the prepared state
and the quality of the hardware samples which should have sufficient overlap with the target state for any practical advantage against classical methods.
For this, chemistry-inspired hardware-efficient ansatz can be used with parameters optimized in a sub-optimal manner via VQE.\cite{motta2023bridging, motta2024quantum} However, such an approach
adds a heuristic component to the workflow which does not guarantee a substantial overlap with the target state to show
any deterministic practical advantage.
To this end, recently active research is going on to enhance the performance of SQD to push its limits toward better sampling of the
configurations\cite{piccinelli2025quantum, danilov2025enhancing}.
Given that the state fidelity and overall performance of the SQD pipeline is directly dependent on the quality of this recovery step, the development of a more
robust and efficient
configuration recovery protocol with better convergence guarantees is critical for the optimal utilization of HPC resources and the scalability of QCSC algorithms as a whole.

In this work, we propose a QCSC workflow driven by a physics-informed generative machine learning (ML) embedded
configuration recovery strategy termed PIGen-SQD
for accurate fermionic state reconstruction toward improving the efficiency and scalability of SQD algorithm.
The development has two primary objectives. First, it provides a physics-guided, deterministic support to the target state via perturbative measures that complement the heuristic state-preparation step, thereby ensuring a substantial initial overlap for the configuration-recovery module. Second, it leverages the ability of generative ML models to learn complex nonlinear distributions in order to reduce the effective diagonalization dimension through classical sampling from the dominant sector of the Hilbert space. To facilitate this generative ML-based configuration recovery, restricted Boltzmann machines (RBM) are used in this work
which shows immense promise to solve many-fermionic problems\cite{herzog2023solving, xia2018quantum, singh2025maximal, halder2024machine,halder2025construction, hernandez2025configuration}.
Our numerical demonstrations on strongly correlated molecular systems indicate that PIGen-SQD reduces the diagonalization subspace dimension by up to
70$\%$ relative to standard SQD, while simultaneously yielding energies that are orders-of-magnitude more accurate. This highlights its ability to reconstruct fermionic states efficiently by exploring only the chemically relevant sector of the Hilbert space. Such a strategic orchestration of classical many-body theory, quantum sampling, and the generative capabilities of machine learning results in an QCSC framework that is accurate, efficient, and systematically improvable.

The paper is structured as follows:
we start with a brief review of QCSC and many-body perturbation theory.
Next, we discuss the efficient computation of perturbative estimates for the target-state support and outline the details of the RBM model.
Finally, we present numerical results of PIGen-SQD against SQD and other standard electronic structure methods to demonstrate the superior performance of our method.

\section{Brief Overview of Quantum Centric Supercomputing} \label{QCSC}

The primary goal of quantum chemistry is to find the eigenvalues and eigen vectors of the many-fermionic molecular Hamiltonian
\begin{equation}
    \hat{H} = \sum_{pq} h_{pq} \hat{a}^{\dagger}_p \hat{a}_q + \sum_{pqrs} v_{pqrs} \hat{a}^{\dagger}_p \hat{a}^{\dagger}_q \hat{a}_s \hat{a}_r
\end{equation}
where, $p,q,r,s$ are molecular orbital indices, $h_{pq}$ and $v_{pqrs}$ are respectively the one and two electron integrals in the molecular orbital basis
and $\hat{a}_p,\hat{a}^{\dagger}_q$ are the creation and annihilation operators. 
Under Born-Oppenheimer approximation, the corresponding exact solution can be expressed as a linear combination of many-body basis formed by a reference state $\ket{\Phi_0}$ generally taken to be Hartree-Fock (HF) determinant and a set of excited determinants (with respect to HF)
\begin{equation} \label{CI expansion}
    \ket{\Psi} = \ket{\Phi_0} + \sum_{ia} c_{i}^{a} \ket{\Phi_{i}^{a}} + \sum_{\substack{ijab \\ j>i,\, b>a}} c_{ij}^{ab} \ket{\Phi_{ij}^{ab}} + ..
\end{equation}
where, the set $\{i,j,k,l,m..\}$ ($\{a,b,c,d,e..\}$) represents occupied (virtual) spin-orbitals in the HF picture in a given molecular orbital basis.
If the solution in Eq. \eqref{CI expansion} contains all the way up to $N_e$-th rank excitations (where $N_e$ is the total number of particles),
the resulting wavefunction corresponds to a full configuration interaction (FCI) expansion. Solving an FCI problem is equivalent to exact diagonalization of the Hamiltonian
in the given many body basis.
For a molecular problem with $M$ spin orbitals,
the states in the entire $2^M$ dimensional Hilbert space
which violate fundamental symmetries such as the number of electrons or the total spin, have no contribution to the target state.  
Thus, the target state $\ket{\Psi}$ can be expressed as a linear superposition of only the symmetry-preserving subset $Q$ of the entire Hilbert space
\begin{equation} \label{symmetry space ci expansion}
    \ket{\Psi} = \sum_{\mu \in Q} c_{\mu} \ket{\Phi_{\mu}}
\end{equation}
where $\mu$ are the composite hole-particle indices.
The dimension of $Q$ space $d_Q$ for closed-shell systems is given by 
\begin{equation} \label{eq d_Q symmetry space}
    d_Q = \binom{M/2}{N_{\alpha}}.\binom{M/2}{N_{\beta}}
\end{equation}
with $N_{\alpha}$ ($N_{\beta}$)
being the number of electrons in the up (down) spin sectors. 
Generally the FCI coefficients $c_\mu$ for some low-lying eigen states are obtained via projection
and diagonalization of the Hamiltonian operator in the symmetry space using efficient diagonalization methods like Davidson's algorithm\cite{DAVIDSON197587}.
However, since the dimension of the symmetry space grows combinatorially with the system size, FCI method cannot be applied to practically relevant systems.
Usually $Q$ space is sparse since a relatively huge number of configurations have zero coefficients and are redundant in the CI expansion Eq. \eqref{symmetry space ci expansion}.
The subspace which contains determinants with only non-zero coefficients is to be referred to as the \textit{core space}
$C$ with number of determinants $d_C$ which is in practice much smaller than $d_Q$.
Finding out the exact core space is an extremely non-trivial task and the holy grail of any CI method as
the prior information of the core space drastically reduces
the dimensionality of the diagonalization task for better scalability.
Toward this, several selected CI methods have been developed recently for selecting subspaces that approximate core space
while retaining the desired accuracy\cite{sherrill1999configuration,holmes2016heat, liu2016ici, sharma2017semistochastic, hu2024small}.

Quantum centric supercomputing (QCSC) or quantum selected CI (QSCI) methods are a class of hybrid quantum-classical algorithms
designed to outsource this classically nontrivial core space selection task
to a quantum processor. This is followed by projection and diagonalization of the corresponding Hamiltonian in the selected space on a classical computer for energy estimation.
Toward this, the fermionic electronic-structure problem is mapped to the quantum-computing framework by encoding the M molecular spin-orbitals onto
M qubits, such that each electronic configuration corresponds to a computational-basis state $\ket{x}$ where ${x} \in \{0,1\}^M$.
The single-qubit states $\ket{0}$ or $ \ket{1}$ represents unoccupied and occupied spin-orbitals, respectively.
Jordan-Wigner transformation is used to map the corresponding second-quantized fermionic operators to qubit operators.
In the QCSC framework, initially a trial state $\ket{\Psi}$ on $M$ qubits (which represent $M$ molecular spin orbitals)
is prepared on a quantum computer with a parameterized quantum circuit (PQC)
that should have overlap with the true target state. This is ensured by preparing a state with a physically motivated parameterized ansatz structure
of the general form
\begin{equation}
    \ket{\Psi} = e^{\hat{G}} \ket{\Phi_0}
\end{equation}
where the unitary operator $e^{\hat{G}}$ (ensured by the anti-Hermiticity of the operator $\hat{G}$) explicitly introduces electronic corrleation
by acting upon the reference state.
This PQC is executed on a quantum computer to perform repeated measurements in the computational basis
that produce a distribution of measurement outcomes
\begin{equation}
    \tilde{\chi} = \{ x | x \sim \tilde{P}_{\Psi}(x) \}
\end{equation}
where $\boldsymbol{x} \in \{0,1\}^M$ is computational basis states (or slater determinants)
along with the associated noisy probability distribution $\tilde{P}_{\Psi}(x)$ obtained from the sampled states according to
\begin{equation}
    \tilde{P}_{\Psi}(x) = \frac{N_{\boldsymbol{x}}}{N_{shots}}
\end{equation}
Here, $N_{\boldsymbol{x}}$ is the frequency of the configuration $\boldsymbol{x}$ in the sample distribution and $N_{shots}$ is the total number of shots.
In an ideal noiseless scenario the sampled distribution contains the dominant electronic configurations from symmetry space $Q$ according to the noiseless distribution $P_{\Psi(x)}$ with the reference configuration
being the most frequent. One then diagonalizes the Hamiltonian in the sampled subspace which results in the upper bound energy estimation of the
target state. However, in presence of hardware noise, several spurious states contaminate the sampled space with particle or spin non-conserving determinants.
This warrants the need for a robust post-processing technique in the QCSC framework that discards the spurious states and recovers the symmetry-preserving configurations in the diagonalization space based on the information from hardware samples,
along with exploring additional dominant configurations that the quantum samples might have missed due to hardware noise and inadequate number of shots.
The recently developed SQD\cite{robledo2025chemistry} method introduces a self-consistent \textit{configuration recovery} protocol
to restore the dominant configurations. SQD uses the local unitary coupled jastrow (LUCJ)\cite{motta2023bridging} ansatz which is a variant of the unitary coupled jastrow (UCJ)\cite{matsuzawa2020jastrow}
with a \enquote{local} approximation to respect the limited qubit connectivity of a quantum hardware while
preserving a degree of physical intuition of the corresponding molecular system.
LUCJ with a single layer has the form
\begin{equation} \label{LUCJ}
    \ket{\Psi}_{LUCJ} = e^{-\hat{K}_2} e^{\hat{K}_1} e^{i\hat{J}_1} e^{-\hat{K}_1} \ket{\Phi_0}
\end{equation}
where, $\hat{K}_1$ and $\hat{K}_2$ are one-body operators and $\hat{J}_1=\sum_{p\sigma r \tau}J_{p\sigma,r\tau}\hat{n}_{p\sigma}\hat{n}_{r\tau}$
is a density-density operators with $\hat{n}$ being the number operator.
In practice, the LUCJ parameters can be initialized with jastrow factorization of the coupled cluster with singles and doubles (CCSD) amplitudes
and can be optimized further variationally for a better state preparation\cite{motta2024quantum}.
With the hardware samples obtained from quantum circuits with LUCJ approximation, SQD configuration recovery module
filters out the spin and particle number symmetry preserving state and partially recovers
a new set of configurations $\chi_R$ from noisy samples $\tilde{\chi}$ with a problem-motivated clustering of configurations using average orbital occupations. 
This is followed by an SCI loop for Hamiltonian projection and diagonalization 
performed in $K$ batches each containing $d$ configurations
sampled from $\chi_R$. While the SQD workflow produces accurate results, it is sensitive to the quality of the hardware samples.
The presence of huge number of spurious
symmetry non-preserving configuration is extremely likely and it requires substantially large number of samples to draw and search for CR
module to effectively find out dominant configurations which lie within the core space. This ultimately proliferates the diagonalization dimension
which could be the main bottleneck for scalability using HPC resources.
To address this issue, in this work, we propose a novel QCSC strategy to restrict the search space and diagonalization dimension of the CR module by providing judicious support 
to the hardware samples by first-principle many-body perturbative estimates with low rank implicit tensor decompositions and subsequent guided subspace expansion with generative ML.


\section{Efficient Configuration Recovery via Physics-Informed Modeling: Perturbative Estimation for Configuration Selection}

In the presence of hardware noise and a finite number of shots, even chemistry-inspired hardware-efficient ansatze struggle to generate a sufficient number of samples within the core space\cite{robledo2025chemistry}
which ultimately determine the accuracy of chemically relevant observables. Although such noisy samples can still serve as input for configuration recovery, subsequently this step often requires diagonalizing a large subspace in order to isolate the true core space configurations. This diagonalization constitutes one of the primary bottlenecks of QCSC/QSCI workflows and significantly limits scalability.
To alleviate this bottleneck, we introduce a first-principles strategy that reliably identifies a dominant configuration subspace from the core space.
This subspace, when combined with the symmetry-preserving hardware samples,
results in a physics-informed distribution concentrated on configurations with substantial overlap with the target state.
Such a distribution provides an excellent initialization for any self-consistent configuration-recovery module, enhancing both the efficiency and accuracy of subsequent fermionic state reconstruction toward better scalability.



\subsection{Anchoring the Hardware Samples to Ground State via Many-Body Theories} \label{MBPT theory}

Starting from the first principle many-body 
theory, an unbiased ensemble of the ground state configurations may be built up
via M{\o}ller-Plesset perturbation theory (MPPT)\cite{moller1934note,cremer2011moller} with the perturbing potential
\begin{equation}
    \hat{V} = \sum_{i}\sum_{j>i} \frac{1}{r_{ij}} - \sum_i v_{HF}(i)
\end{equation}
where $v_{HF}$ denotes the Hartree-Fock 
potential.
The corresponding Hilbert space is split into a model space spanned by a reference function $\ket{\Phi_0}$ 
and a perturbative orthogonal space $\ket{\chi}$ spanned by excited determinants defined by
\begin{equation}
    \ket{\Phi_{ijk...}^{abc...}} = \hat{a}^{\dagger}_a \hat{a}^{\dagger}_b \hat{a}^{\dagger}_c ....\hat{a}_k \hat{a}_j \hat{a}_i \ket{\Phi_0}.
\end{equation}
In the perturbative expansion, the exact ground state $\ket{\Psi}$ can be written as
\begin{equation}
    \ket{\Psi} = \ket{\Phi_{0}} + \ket{\chi} = \ket{\Phi_{0}} + \lambda \ket{\Psi^{(1)}} + \lambda^2 \ket{\Psi^{(2)}} + ..
\end{equation}
where, $\lambda$ is the order parameter and $\ket{\Psi^{(n)}}$ is the $n$-th order perturbative correction to the wavefunction.
The Slater-Condon rule\cite{szabo1996modern} ensures that the first order correction involves only doubly excited determinants\cite{shavitt2009many}
\begin{equation}
    \ket{\Psi^{(1)}} = 
    \sum_{\substack{ijab \\ j>i,\, b>a}} 
    v_{ij}^{ab}\Big({\Delta_{ij}^{ab}}\Big)^{-1} 
    \ket{\Phi_{ij}^{ab}} = \sum_{\substack{ijab \\ j>i,\, b>a}} 
    t_{ij}^{ab} 
    \ket{\Phi_{ij}^{ab}}
\end{equation}
where, $v_{ij}^{ab}=\langle ij ||ab\rangle = \langle ij |ab\rangle - \langle ij |ba\rangle$ with quantities of $\langle ... |...\rangle$ type being the two-electron integrals,
$t_{ij}^{ab}$ are the second order M{\o}ller-Plesset (MP2) amplitudes that are generated as a part of the HF module and $\Delta_{ij}^{ab}$ are the MP2
denominators.\cite{cremer2011moller} In the first order correction to the wavefunction, only those doubly excited determinants contribute for which $t_{ij}^{ab}$ is
non-zero or in practice,
higher than a pre-defined threshold $\epsilon_{int}$.
The same numerical threshold $\epsilon_{int}$ would be used to filter out all the subsequent integrals as well.

\subsection*{Efficient Generation of Higher-rank Perturbative Terms via Low-Rank Tensor Decomposition and Symbolic Algebraic Operations:} \label{efficient higher rank generation and judicious index manipulation}

Strong correlation often warrants the inclusion of high rank excitations for better representation of the wavefunction. In view of this, we further provide a better support to the ground state wavefunction by identifying the configurations that appear in the second and higher order perturbative rungs. A hierarchical configuration
screening ensures that some of the most dominant high-rank
configurations are selected to supplement the hardware samples.
For the selection of any dominant configuration at a given perturbative order, we analyze in terms of the excitation structure out of the HF vacuum. For any excitation structure (that generates an excited configuration by their action of the vacuum), the 
many-body perturbation theory ensures that there exists an excitation structure of immediate lower rank which has
dominant contribution to the wavefunction. This translates to the fact that the selection of any 
higher rank excited configuration may be performed by symbolic (rather than numeric) manipulation 
(or equivalently excitation structure identification) with the knowledge of immediate lower rank configurations.

When stepping from the first- to the second-order perturbative correction of the wavefunction, the correlation space is spanned by single, double, and triple excitations. Since the doubly excited configurations have already been filtered during the first-order correction, in the second-order correction to the wavefunction, we focus only on the selection of the dominant singly and triply excited configurations 
\begin{equation}
\begin{split}
    & \ket{\Psi^{(2)}} \leftarrow \sum_{\substack{ia \\ }} c_i^a \ket{\Phi_{i}^{a}} + \sum_{\substack{ijkabc \\ k>j>i\\ c>b>a }} c_{ijk}^{abc} \ket{\Phi_{ijk}^{abc}} \\
\end{split}
\end{equation}
where $c_i^a$, $c_{ij}^{ab}$ are the expansion coefficients of the singles and triples.
From the  MPPT calculations, approximate measures of these coefficients in the many-body basis may be represented as\cite{halder2024noise}:
\begin{equation} \label{c_ia c_ijab def}
\begin{split}
    & c_i^a \leftarrow \Big({\Delta_{i}^{a}}\Big)^{-1} \Big[ \sum_{\substack{m,n \in occ \\ e \in virt}} v_{ie}^{mn} t_{mn}^{ae} + \sum_{\substack{m \in occ \\ e, f \in virt}} v_{ef}^{am} t_{im}^{ef}  \Big] \\
    & c_{ijk}^{abc} \leftarrow \Big({\Delta_{ijk}^{abc}}\Big)^{-1} \Big[ \sum_{m \in occ} v_{ij}^{am} t_{mk}^{bc} + \sum_{e \in virt} v_{ie}^{ab} t_{jk}^{ec} \Big].
\end{split}
\end{equation}
Owing to the underlying perturbative structure as discussed previously, only those rank-three excited configurations are retained in the first place which subsumes the already selected rank-two configurations, or in other words, the triply excited configurations must be reachable from the rank-two excited configurations by \textit{scattering} structures like $a_a^\dagger a_m^\dagger a_ja_i$ or $a_a^\dagger a_b^\dagger a_e a_i$. These are rank-two operators with effective excitation rank of one and are referred to as \textit{scatterers}, $\hat{S}$. The associated integrals (their first order measure) for $\hat{S}$ are given as $v_{ij}^{am}$ and $v_{ie}^{ab}$. 
The coefficient of such a triply excited configuration ($c_{\mu_3}$) generated by 
an excitation structure like $\mu_3$ (where the composite hole-particle index $\mu_3 \equiv \{ijkabc\}$ with k>j>i and c>b>a) at the second
perturbative order may be ascertained by the tensor contraction between the excitation operators (which is already selected at the first order) and scatterers,
\begin{equation} \label{cmu3}
    c_{\mu_3} \leftarrow \Big| \Big[ (\hat{S}_{I}  \hat{T}_{\mu_2})_c \Big]_{\mu_3} \Big| \hspace{2mm} \forall \mu_3
\end{equation}
where the braces $(..)_c$ denote tensor contractions via common contractible 
hole or particle indices in the associated operators ($\hat{S}_I$ and $\hat{T}_{\mu_2}$ in this case) and $\mu_n$ denotes the composite hole-particle
indices for $n$-th rank excitation.
An ideal measure of the first order importance of the selected triple excitations would
be to perform the explicit
summation over the contractible indices. However, this would 
entail significant computational overhead and is generally impractical.
A more viable alternative is to choose only those $\mu_3$ if there exists a $v_I>\epsilon_{int}$ with 
the scattering index $I$ sharing
contractible indices with the underlying two-body excitation structure $\mu_2$.
Given $N_s$ and $N_d$ number of  elements present in pruned $\hat{S}_I$ and $\hat{T}_{\mu_2}$,
such an operation would require $\mathcal{O}(N_sN_d)$ number of operations at worst.

Nonetheless, since our subsequent workflow does not
necessitate such a stringent perturbative measure beyond ensuring adequate support to the target state,
we employ an efficient strategy here involving judicious index handling.
With the pruned set of $\hat{S}_I$ and $\hat{T}_{\mu_2}$ operators, ideally
one needs to perform all the tensor contractions of the terms given in Eq. \eqref{tensor contraction terms for triples} (see Appendix \ref{PT algebraic terms}).
Among these terms, computation of $  v_{ie}^{ab} t_{jk}^{ec} \rightarrow c_{ijk}^{abc}$ is the most expensive as a set of contractible index $e$ belongs to the virtual orbital indices.
However, due to the sparsity of two body integrals and MP2 amplitudes, this term can be efficiently calculated using a
symbolic partial tensor contraction technique for minimal number of operations.
Since only the outer indices (non-contractible) post tensor contraction are of importance as the triples are ultimately governed by these outer indices,
to reduce the number of operations stemming from explicit tensor contractions, only those combination of outer indices, that originate
from pruned $I$ and $\mu_2$ sets, are retained which share only one common index from the contractible set of indices (see Appendix \ref{appendix symbolic tensor contraction}).
Being a symbolic operation, this would not require explicit summation over the entire range of a particular contractible index and the associated cost is in practice much less than $\mathcal{O}(N_sN_d)$.
Such index handling provides the flexibility to avoid the prohibitive cost of explicit tensor contractions, saving huge number of operations to be performed
which is further discussed quantitatively in Sec.\ref{computational details}.

Generalizing this approach for higher order terms, the coefficients for the $n$-tuply excited configurations can be obtained by evaluating the associated amplitudes of term like
\begin{equation} \label{c_mu_n}
    c_{\mu_n} \leftarrow \Big| \Big[ \Big((((\hat{S}_{I_1} \hat{S}_{I_2})_c\hat{S}_{I_3})_c...)_c\hat{S}_{I_{n-2}})_c  \hat{T}_{\mu_2} \Big)_c \Big]_{\mu_n} \Big| \hspace{2mm} \forall \mu_n
\end{equation}
At each order of expansion we perform a hierarchical pruning as described above.
Since only a small ensemble of configurations contribute significantly to 
the true wavefunction, these perturbative calculations allow one to 
efficiently select a substantial amount of the important determinants
which can be systematically improved by increasing $n$ (along with the associated computational cost).
Our numerical analysis shows that the importance selection for configurations 
obtained from perturbative rank up to $n=4$ provides desired support for 
the subsequent generative learning which requires asymptotically $\mathcal{O}(N_s^2N_d)$ operations at worst for configuration screening. However, decomposing the corresponding terms (see Appendix \ref{PT algebraic terms}) into intermediates along with the symbolic operation discussed so far, the associated cost would be much reduced. Upon adding the perturbatively obtained
configurations to the filtered symmetry-preserved hardware samples, we get a much 
better distribution of dominant determinants which can be used as an initial point 
for any self-consistent subspace generation method to predict more determinants 
from the core space for energy estimation. Toward this, in the next 
section we review the inner workings of a generative ML method that efficiently 
scans \textit{only} an important sector of the symmetry-preserved Hilbert space 
and further generates important configurations for diagonalization in a 
self-consistent manner.

\subsection{Generative Machine Learning Based Fermionic State Reconstruction} \label{ML based configuration recovery}

Restricted Boltzmann Machine (RBM) is a generative stochastic unsupervised machine learning model that has the ability to learn
complex nonlinear probability distributions
through a bipartite graphical structure consisting of a
layer of visible $\boldsymbol{g} = (\{ g_i\}_{i=1}^D, g_i \in\{0,1\})$ and hidden
$\boldsymbol{h} = (\{ h_j\}_{j=1}^J, h_j \in\{0,1\})$ units.
The visible layer encodes the state of the corresponding system and the hidden variables introduce additional
degree of freedom for enhanced expressivity.
With the training data taken as the input vector for visible layer, RBMs optimize a set of 
parameters $\theta=\{\boldsymbol{u,y,W}\}$ to learn and accurately model the underlying distribution where, $\boldsymbol{u,y}$ are bias vectors for
visible and hidden layers and $\boldsymbol{W}$ is the weight matrix that connects visible and hidden units. From a mathematical point of view, RBMs represent a
joint probability distribution over the visible and hidden layers governed by an energy function
\begin{equation}
    E(\boldsymbol{g,h},\theta) = -\sum_{i=1}^D \sum_{j=1}^Jg_i W_{ij}h_j - \sum_{j=1}^J {y}_j h_j - \sum_{i=1}^D u_i g_i .
\end{equation}
The associated probability distribution of RBM follows the Gibbs distribution 
\begin{equation}
    P(\boldsymbol{g};\theta) = \sum_h P(\boldsymbol{g,h}, \theta) = \sum_h \frac{1}{Z(\theta)} exp(-E(\boldsymbol{g,h},\theta))    
\end{equation}
where,
\begin{equation}
    Z(\theta)= \sum_h \sum_v exp(-E(\boldsymbol{g,h},\theta))
\end{equation}
is the partition function.
Since explicit calculation of the partition function becomes computationally intractable,
RBMs are usually trained on a set of vectors $\boldsymbol{g}$ distributed according to some unknown probability distribution
using the contrastive divergence (CD)\cite{hinton2012practical} method which approximates the gradient of the log likelihood
to learn the underlying distribution. RBM alternatively updates hidden
and visible layers using conditional activation probabilities:
\begin{equation}
    p(h_j=1|\boldsymbol{g}) = \sigma\Big( \sum_i g_i W_{ij} + y_j \Big)
\end{equation}
\begin{equation}
    p(v_i=1|\boldsymbol{h}) = \sigma\Big( \sum_j W_{ij} h_j  + u_i \Big)
\end{equation}
where, $\sigma(x)=\frac{1}{1+e^{-x}}$ is the sigmoid function.
By maximizing the log likelihood of the data $\ln P_{\theta}(\boldsymbol{g})$,
RBM uses stochastic approximation to update its parameters which is equivalent to minimizing the Kullback-Leibler
divergence between the data distribution and model distribution.
After training, a new set of vectors are generated according to the learnt distribution $P(\boldsymbol{g})$ using Gibbs sampling\cite{mackay2003information} technique.
For a detailed description of the working principle of RBM, the readers are referred to Refs.\cite{hinton2012practical,zhang2018overview, hinton2002training, hinton2006reducing, melko2019restricted}.

In the context of electronic structure theory, the representational capacity of the RBM is particularly relevant, as it can efficiently capture intricate correlations among electronic configurations\cite{herzog2023solving, xia2018quantum, singh2025maximal, halder2024machine,halder2025construction, hernandez2025configuration}. A properly trained RBM guides the sampling toward regions of the Hilbert space dominated by determinants with the largest contributions to the CI wave function, thereby markedly reducing the dimension of the diagonalization subspace and
the associated computational expense.
However, being an unsupervised learning model, the training of RBM is sensitive to the quality of the input vector (visible layer)
which is used to predict the true probability distribution.
In some recent works\cite{herzog2023solving,hernandez2025configuration}, dominant configurations from a CISD calculation is used
as the starting input vector for training. However, since CISD can qualitatively fail to describe systems with strong electronic correlation, there remains
ample room for constructing better physics-informed training datasets that more faithfully represent the true distribution of electronic configurations.
This can lead to more accurate predictions and a more focused exploration of the configuration space.
Toward this, a more uniformly distributed collection of dominant determinants, sampled across different sectors of the symmetry-preserving Hilbert space, would yield a significantly more balanced training dataset for the RBM.
Leveraging this generative capability, the present work explores the role of RBMs in enhancing the configuration-recovery module within the SQD framework, wherein hardware-generated samples are supplemented by perturbatively predicted dominant states to enable physics-informed guided learning.
We refer this approach as physics-informed generative machine learning based SQD or PIGen-SQD.


The workflow starts with perturbative calculations described in section \ref{MBPT theory}.
Once the perturbative selection is done, it can be appended with the
sampled determinants filtered to contain only the correct particle sector configurations.
The molecular Hamiltonian is then diagonalized in this computational basis which provides a set of CI coefficients $c_{\mu_n}$ where $\mu_n$ is a general
coefficient of $n$-tuply excited determinant with $0\leq n \leq N_e$ ($N_e$ being the total number of electrons).
Upon diagonalization, the states with coefficients greater than a threshold $\epsilon_{coeff}$ is used to prepare the input vector of the RBM according to their CI coefficient weights while the ones with less than $\epsilon_{coeff}$
(which are practically zero) coefficients are identified and are stored in a \enquote{blacklist} to avoid re-selection.
In this context, it is important to note that in the presence of only perturbatively determined $n$-tuply excited configurations,
all determinants with excitation rank 
$>(n+2)$ would, by the Slater-Condon rules\cite{szabo1996modern}, be assigned to the blacklist and consequently
excluded from the diagonalization subspace.
This would result in systematically missing important higher-order excitations and lead to unreliable configuration recovery,
particularly for small values of $n$.
Addressing this issue would necessitate more complex routines and incur additional computational cost.
Here, hardware samples play a crucial role: they are far more likely to include relevant higher-order excitations, thereby mitigating the risk of inadvertently excluding important configurations.
The prepared input vector containing hardware samples and perturbatively selected configurations is used for generative-ML training and subsequent state generation, which is
performed iteratively till it reaches the convergence criteria. The entire algorithm 
requires the following sequential steps:


\begin{enumerate}
    \item \textbf{Initial State Preparation and Sampling from Quantum Hardware:} State preparation is done on a quantum hardware with an ansatz that has a substantial overlap with the ground state. Usually a PQC corresponding to a unitary $\hat{U}(\theta)$
    is prepared that acts on HF state $\ket{\Phi_0}$ such that the prepared state is $\ket{\Psi} = \hat{U}(\theta) \ket{\Phi_0}$. This is followed by repeated measurements performed with a specific number of shots $N_{shots}$ to obtain samples from the hardware that produce a measurement outcome $\tilde{\chi} = \{ x | x \sim \tilde{P}_{\Psi}(x) \}$ with a probability distribution $\tilde{P}_{\Psi}(x)$ where each bitstring $x \in \{0,1\}^M$ is a computational basis state. 

    
    \item \textbf{Restoration of Fundamental Symmetry and Physics-Informed Initialization to Configuration Recovery:} Symmetry preserved configurations are filtered from the noisy distribution and retained for post processing after normalization of the state vector. Perturbative calculations are performed using the index-based method described in section \ref{MBPT theory}. This step provides a set of dominant configurations
    of rank up to $n$ chosen by the user.
    The filtered hardware configurations and perturbatively obtained configurations are stacked together
    to provide a substantial support to the sampled distribution for a physics-informed initialization for configuration recovery.



    
    \item \textbf{Fermionic State Reconstruction through Generative ML-based Self-Consistent Configuration Recovery:}
    Configuration recovery is performed with RBM in alternating micro- and macro-cycles in the following manner
    
    \textbf{Macro Cycle:} In the $k$-th iteration of macro-cycle, dominant configurations obtained from $(k-1)$-th step is used for subspace projection and diagonalization of the
    molecular Hamiltonian and the corresponding eigen vectors (CI coefficients $c_{\mu_n}$) and eigen values are obtained. Post diagonalization, a user-defined threshold $\epsilon_{coeff}$ is set to filter only those computational basis states for which $c_{\mu_n}>\epsilon_{coeff}$. The coefficients with
    $c_{\mu_n}<\epsilon_{coeff}$ are stored in an array and labeled as \enquote{blacklisted} configurations since they do not have major contribution to the state. These configurations will be avoided
    in further diagonalizations even if they are selected as a part of the configuration recovery module. The filtered dominant configurations are processed to construct the sample distribution for ML training where the frequency of each sample configuration is proportional to its CI coefficient obtained by diagonalization. However, HF configuration, which typically has a large CI coefficient and consequently very high probability, is discarded from the training set as it can bias the learning. Post new sample generation, it is again added to the configuration list for diagonalization. 

    

    
    \textbf{Micro Cycle:} Each macro-cycle contains a set of micro-cycle iterations which is used to train RBM with the prepared input vector distribution. Trained RBM is used to generate new configurations using Gibbs sampling method. Toward this, a set of spin-particle conserved configurations, whose number is fixed at $x$\% of $d_Q$, are randomly sampled from the symmetry space and given to the Gibbs sampler which generate same number of spin-particle symmetry conserved configurations.



    \item \textbf{Recursive Subspace Expansion through Hamiltonian Diagonalization:} The generated configurations are compared against the blacklisted configurations to avoid re-selection. After removing any blacklisted configurations present in the newly generated distribution, it is combined with the existing
    dominant configurations for a new expanded subspace. This subspace is used again in $(k+1)$-th macro-cycle and this alternate macro and micro-cycles are repeated until there is no further generation of new states or the
    energy between two successive macro-iterations is less than a user-defined threshold $\epsilon_E$.

    


\end{enumerate}

The synergy between generative machine learning models such as the RBM and perturbative insights from calculations discussed in section \ref{MBPT theory}
enables a substantially more efficient and scalable implementation of the SQD configuration recovery workflow.
In the next section we will show the numerical efficiency and accuracy of the MBPT-RBM driven configuration protocol compared to the standard SQD method.

\begin{figure}[!ht]
    \centering
\includegraphics[width=\linewidth]{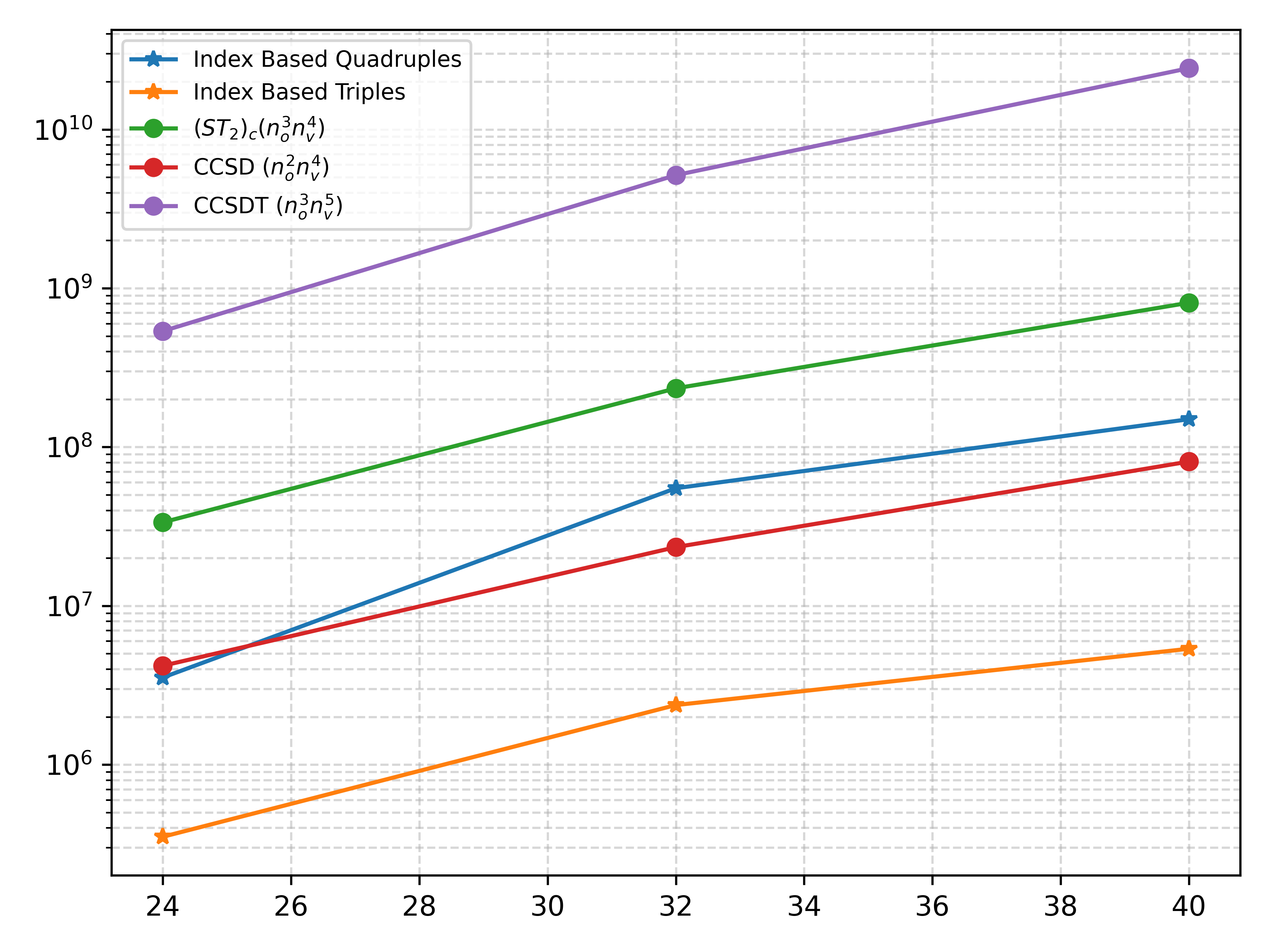}
\caption{\textbf{Cost comparison between numerical worst case scenarios for perturbative triples and quadruples selection and the corresponding asymptotic scaling of the term $v_{ie}^{ab} t_{jk}^{ec}$
with respect to the number of qubits. Cost of CCSD and CCSDT is also plotted for reference. Here $n_o$ and $n_V$ are the number of occupied and virtual spin orbitals respectively. Our analysis shows the maximum cost for triples is almost two orders-of-magnitude less compared to the asymptotic scaling making it an efficient method for approximate perturbative selection of excited configurations. The corresponding quadruples selection cost scales almost similar to CCSD (or marginally worse).} }
    \label{st contraction cost comparison}
\end{figure}


\section{Computational Details} \label{computational details}
All the numerical results in this work are obtained by in-house python based codes. MBPT calculations are done using two-body integrals and MP2
amplitudes obtained from PySCF\cite{sun2018pyscf} driver in Qiskit\cite{qiskit2024}. Three strongly correlated molecular systems, 
$\mathrm{H_2O}$ (equilibrium O-H bond distance $R_{eq}=0.958\mbox{\AA}$, bond angle $\angle \mathrm{H-O-H} = 104.4776^\circ$), $\mathrm{N_2}$ ($R_{eq}=1\mbox{\AA}$) and $\mathrm{C_2H_2}$ ($R_{eq}=1.202\mbox{\AA}$) are studied for different geometries using 6-31G basis set with frozen core approximation involving
24, 32 and 40 spinorbitals correlating 8, 10 and 10 electrons respectively.
The hardware mapping of the corresponding LUCJ ansatz on IBM Heron R2 processor requires 27, 36 and 45 qubits respectively that includes
the qubits associated with Jordan-Wigner mapped spin orbitals and ancilla qubits.
Numerical experiments are also performed using IBM Heron R3 processor on $\mathrm{H_2O}$ in the cc-pVDZ basis, which comprises 46 spin orbitals, that requires a total of 52 qubits.
Details of the quantum hardware layout and the qubit assignments for all the numerical experiments are given in Appendix \ref{appendix: hardware layout and coupling map}.
The FCI, CCSD, CI with singles and doubles (CISD), MP2 and HF calculations were performed using PySCF with the frozen core approximation.
Since FCI calculations are not feasible for $\mathrm{C_2H_2}$ (6-31G basis) and $\mathrm{H_2O}$ (cc-pVDZ basis) under consideration, heat-bath configuration interaction (HCI)\cite{holmes2016heat} calculations were
done using DICE-solver\cite{sharma2017semistochastic} to get reference energy values
for accuracy comparison.
The entire workflow involves multiple independent computational components orchestrated together as discussed below:

\textbf{Quantum state preparation and hardware sample generation:} State preparation on a quantum hardware is done using the LUCJ ansatz (Eq. \eqref{LUCJ}) with CCSD parameters used
for jastrow decomposition. No further optimization of the LUCJ parameters is performed. For 27 qubits experiment 250000 shots were used while 36, 45 and 52 qubit experiments were performed using
a fixed $\mathrm{10^6}$ shots due to limited availability of quantum processor time and shot budget.
Ideally, at least $d_Q$ (see \eqref{eq d_Q symmetry space}) number of shots are required
for an efficient and meaningful sampling task from a quantum hardware. However, in our case only for the study of $\mathrm{H_2O}$ in 6-31G basis ($d_Q=245025$) the number of shots is comparable to the dimension of the symmetry space.
The circuits were created using \textsc{qiskit-addon-sqd}, which uses the \textsc{ffsim}\cite{ffsim} package to generate the LUCJ circuits. For each circuit, we start with some generic linear layout and invoke the \textsc{VF2PostLayout} transpiler pass to find an isomorphic zig-zag layout (which is required for LUCJ circuits) that conforms to the lowest noise profile as obtained from the backend noise information. The LUCJ circuit is placed on this layout using the \textsc{qiskit transpiler}. We executed the circuits on \textit{ibm\_kingston} and \textit{ibm\_boston}, having 156 qubits IBM Heron R2 and R3 processors respectively. The layouts used for the circuits corresponding to each of the molecules are shown in Appendix \ref{appendix: hardware layout and coupling map} (Fig.~\ref{hardware layout} and \ref{hardware layout R3}). 

\textbf{Numerical Worst-Case Scenario for Symbolic Sparse Tensor Operation toward Configuration Selection:} 

To obtain the physics-informed support to ground state via perturbative measures, initially available
molecular integrals (in molecular orbital basis) are filtered from the original sparse integral tensor using a 
threshold $\epsilon_{int}=10^{-10}$.
All surviving integral values, together with their associated 
index combinations, are collected and arranged into an array to construct $\hat{S}$ 
following the integral index pattern ($v_{ij}^{am}, v_{ie}^{ab}$).
The most expensive computational step for triples selection is governed by the term $ v_{ie}^{ab} t_{jk}^{ec}$
which can be performed efficiently using the symbolic index operation technique discussed in Sec.\ref{efficient higher rank generation and judicious index manipulation}.
For this efficient generation
we combine all unique outer indices
$\{iab\}$ and $\{jkc\}$ having only one index common between them (see Appendix \ref{appendix symbolic tensor contraction}) with the constraints $k>j>i$ and $c>b>a$
respectively, stemming from pruned $v_{ie}^{ab}$ and $t_{jk}^{ec}$.
Such an operation does not require explicit summations over contractible indices, and just one common index suffices the
inclusion of the corresponding outer (non-contractible) indices.
At this stage, it is important to note that the hierarchical process of configuration selection through perturbation
theory automatically ensures that the lower order 
terms, double excitations ($t^{**}_{**}$) from the HF reference, are already pruned which implies $N_d << n_o^2n_v^2$.
This provides the flexibility to avoid the prohibitive cost of explicit tensor contraction, saving huge number of operations to be performed.
A comparison of the number of operations required for the pruned elements in the numerical worst case scenario
for triples and quadruples selection
are shown in Fig.\ref{st contraction cost comparison}
against the formal asymptotic scaling of the term $v_{ie}^{ab} t_{jk}^{ec}$, CCSD and CCSDT.
Fig. \ref{st contraction cost comparison} shows
the numerical limiting case $\mathcal{O}(N_sN_d)$ of symbolic tensor manipulation with pruned elements for triples selection
requires almost two orders-of-magnitude less number of operations compared to the asymptotic cost
of explicit $ v_{ie}^{ab} t_{jk}^{ec}$ computations.
The actual implementational cost for the same using the symbolic technique requires much less number of operations than this
in all practical cases.
A similar approach can be followed for constructing perturbative quadruples as well
by breaking the associated contraction into intermediates and performing the index-based symbolic calculations. The algebraic terms
calculated for selection of quadruply excited configurations are given in Appendix \ref{PT algebraic terms}.
The corresponding numerical limiting case for quadruples selection scales almost similar (or marginally higher) to CCSD as demonstrated in Fig.\ref{st contraction cost comparison}. Here, we have considered a selection threshold $\epsilon_{int} =10^{-10}$ which can be
tuned to a larger value to reduce the number of operations further for sparse tensor contractions that would miss some of the configurations (with comparatively
smaller coefficients in the expansion).

While this way of selection of important high-rank excited configurations through their lowest order perturbative measures is not theoretically exact from a conventional many-body viewpoint, we argue that an absolute accuracy (of choosing the important configurations) is not also warranted for; rather, this measure gives us a strong additional support to the hardware generated configuration samples. Dominant determinants that are missed out (due to noise or perturbatively) would eventually get picked up through generative ML. Also, in a similar spirit, any determinant which are spuriously picked up would get dropped (blacklisted) during the iterative fermionic state reconstruction as will be shown in numerical demonstrations.
Owing to this flexibility in the workflow, the symbolic tensor contractions can be restricted to terms with favorable scaling complexity, while still generating a substantial number of dominant configurations of a given rank that exhibit significant overlap with the target state.
A detailed discussion regarding this is presented in Appendix \ref{appendix symbolic tensor contraction}

\begin{figure*}[!ht]
    \centering  
\includegraphics[width=\textwidth]{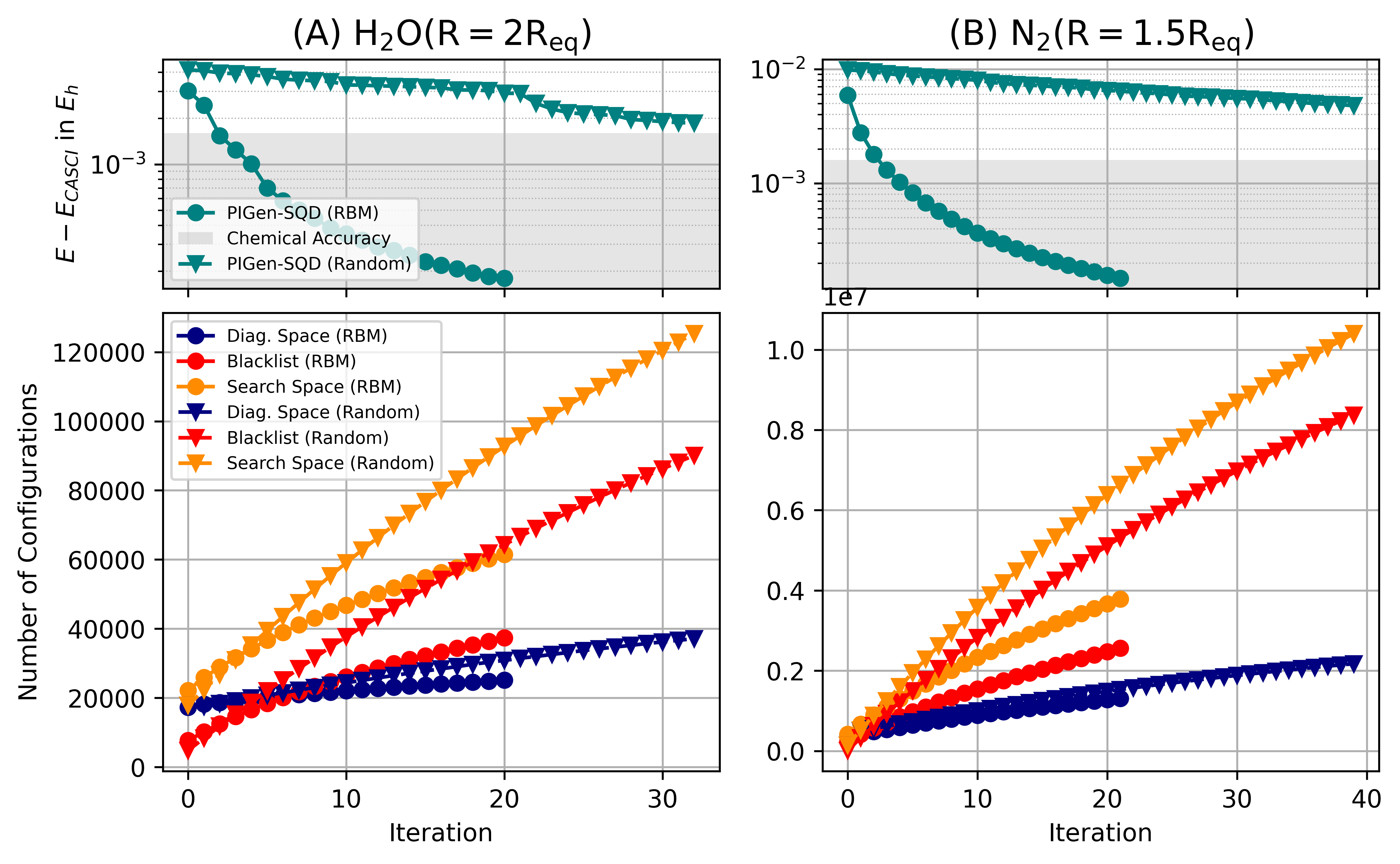}
\caption{\textbf{Convergence profile of PIGen-SQD with RBM driven sampling v/s random sampling from the symmetry space.
The RBM-guided, self-consistent recovery of important configurations leads to noticeably faster convergence and more accurate energy estimates (top row), while simultaneously requiring a significantly smaller diagonalization space and search space (bottom row). This demonstrates that the RBM effectively learns the underlying structure of the dominant sector of the Hilbert space and preferentially samples the most relevant regions, resulting in substantially improved efficiency and accuracy compared to random sampling.
} }
    \label{convergence profile plot}
\end{figure*}

\textbf{Generative ML training and fermionic state reconstruction:}
In this study, we have used RBM with CD algorithm for training.
The dimensions of visible and hidden layer are kept the same in this study which is equal to the number of spin orbitals, the learning rate is fixed at 0.001.
The training distribution provided to the RBM is constructed by inflating the current configuration space obtained in a given macro-iteration, with each configuration replicated in proportion to its associated CI coefficient (excluding the HF configuration, as discussed in Sec.~\ref{ML based configuration recovery}).
Within each RBM micro-cycle, three training iterations are performed using the CD algorithm to update the model parameters. During these updates, CD incorporates an intrinsic Gibbs-sampling step, for which we employ 20 Gibbs iterations.
With $n_{train}$ number of training iterations (or epochs), $n_G$ intrinsic Gibbs sampling steps, $n_{inp}$ inputs, $D$ visible layers and $J$  hidden layers,
the number of operations is bounded by $\mathcal{O}(n_{train}.n_{inp}.n_G.D.J).$\cite{ning2018lcd}
During generation, only one Gibbs sampling step is used that takes a fixed number of randomly selected configurations from the correct spin-particle sector of the
Hilbert space and generates the same number of configurations using the learned parameters.
The number of samples used for Gibbs sampling and subsequent generation at each macro-cycle is fixed a priori by $x\%$ configurations of the symmetry space where the number $x$ is specified for the numerical studies in the result section (Sec. \ref{results and discussion}).
Such a training and generation process avoids over-training and retain a low degree of stochasticity in the model
for better generation of new configurations as discussed further in Sec.\ref{results and discussion}.
The generated determinants are screened against a blacklist containing
$N_{B}$ configurations with a
$\mathcal{O}{(N_{B}log N_{B})}$ complexity while the associated diagonalization scales as $\mathcal{O}(N_{det}^2 M^4)$
with $N_{det}$ determinants in the diagonalization space.\cite{herzog2023solving,davidson197514}
For construction of the blacklist, the associated cutoff threshold is taken as $\epsilon_{coeff}=10^{-10}$
and convergence threshold for the self-consistent loop is $\epsilon_E=10^{-5}$.

\section{Results and Discussion} \label{results and discussion}

In this section, we numerically demonstrate the accuracy and efficiency of PIGen-SQD against SQD.
To assess the performance of our approach we compute the ground state energies of three strongly correlated systems,
$\mathrm{H_2O}$
(equilibrium O-H bond distance $R_{eq}=0.958\mbox{\AA}$, bond angle $\angle \mathrm{H-O-H} = 104.4776^\circ$),
$\mathrm{N_2}$ ($R_{eq}=1\mbox{\AA}$) and $\mathrm{C_2H_2}$ ($R_{eq}=1.202\mbox{\AA}$) for different geometries using 6-31G basis set.
In addition, we consider $\mathrm{H_2O}$ in the cc-pVDZ basis to assess the performance of PIGen-SQD
with SQD with more quantum hardware resources.
For Gibbs sampling process, the number of generated determinants is fixed throughout the iteration process at $x\%$ of the symmetry
space dimension $d_Q$. The value $x=2\%$ is used in all the studies, unless explicitly mentioned otherwise.

\subsection{RBM Driven Sampling v/s Random Sampling}
To illustrate the efficacy of generative machine learning for configuration recovery, Fig. \ref{convergence profile plot}
presents a convergence comparison between RBM-assisted and randomly sampled PIGen-SQD iterations where
the random samples are taken from the symmetry space.
The energy difference with FCI (top row) and the number of configurations
required in the process (bottom row) are plotted along the y-axes with the common x-axis showing the number of iterations.
The grey shaded region represents the chemical accuracy region having energy error less than $1.6mE_h$ with respect to FCI.
Fig.\ref{convergence profile plot} shows
the energy accuracy iteratively gets better for both RBM and random sampling as more and more important configurations are identified
and included into the diagonalization space. Since the role of ML is to generate new states not present in the input data during each macro-iteration cycle, the training process must be designed carefully to avoid overfitting.
Introducing a controlled amount of randomness during training helps prevent the model from becoming too biased toward the input data, which would otherwise limit its ability to generate a sufficiently diverse set of new configurations.
Thus, during our numerical investigations, the number of training iterations for RBM and
the subsequent number of Gibbs sampling are restricted to 3 and 1 respectively.
Such a choice avoids over-training while keeping a low degree of stochasticity in the model that helps
in generation of new states in a guided manner.
This is reflected in significantly faster convergence and more accurate energy estimation 
with RBM driven generation process compared to random sampling.
However, the optimal choice of number of training and sampling steps may be system-dependent and warrants further investigation through explicit hyperparameter optimisation in future work.

\begin{figure*}[!ht]
    \centering  
\includegraphics[width=\textwidth]{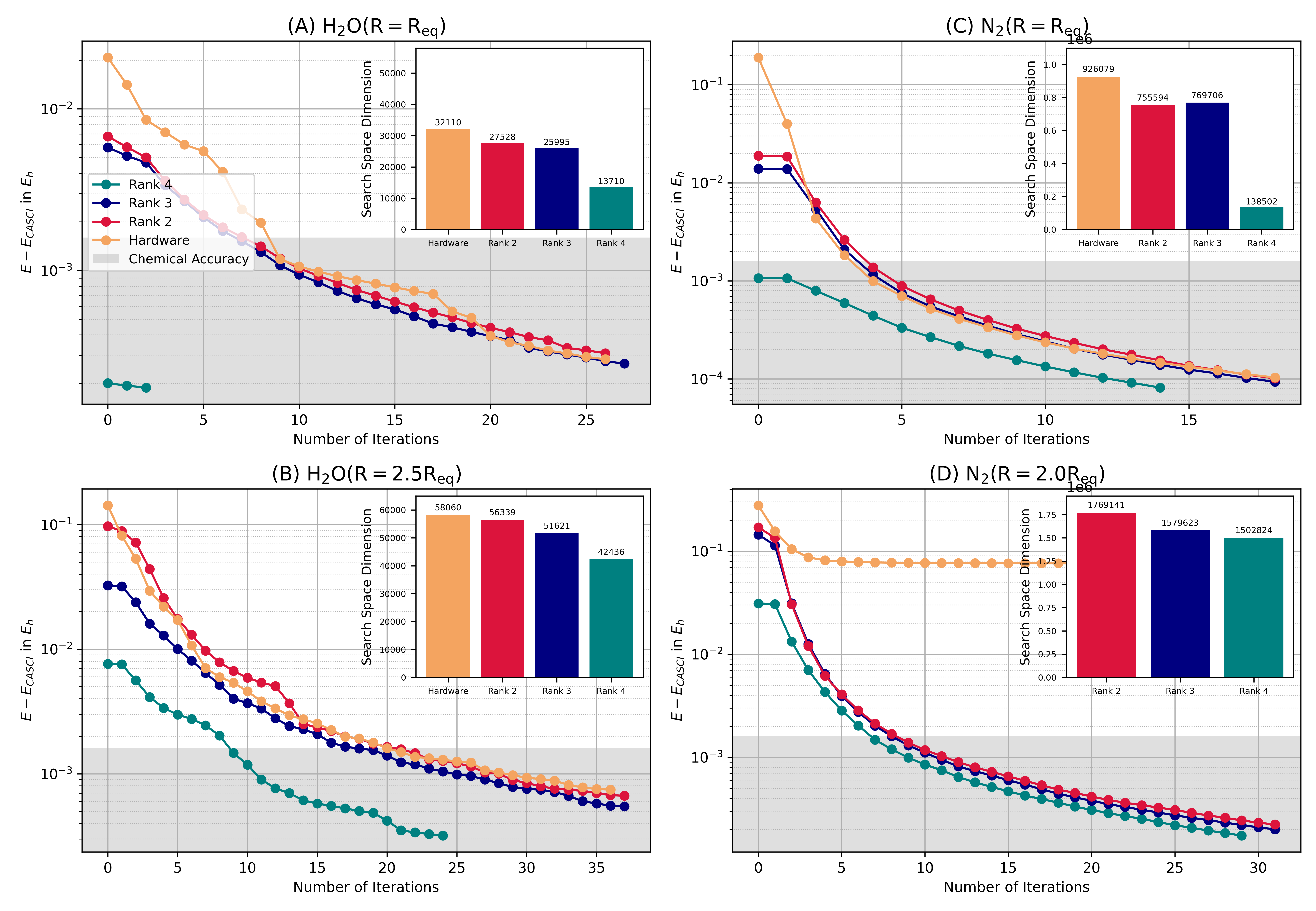}
\caption{\textbf{
Comparison between various PIGen-SQD macro-cycle initialization scheme characterized via different ranks n (n=2,3,4 here) of perturbative estimates and raw (symmetry preserved) hardware samples on the self consistent
configuration recovery process for two different geometries of $\boldsymbol{\mathrm{H_2O}}$ ((A), (B)) and $\boldsymbol{\mathrm{N_2}}$ ((C), (D)). Y-axes show the energy difference compared to FCI with grey shaded regions showing the chemical accuracy. The inset bar plots denote
the dimension of the search space to reach chemical accuracy for each initialization scheme. Quite evidently, initializing the configuration recovery
with perturbative estimates up to rank 4 shows faster convergence with reduced dimensional search space and better energy estimation.} 
}
    \label{mbpt rank comparison}
\end{figure*}

\begin{figure*}[!ht]
    \centering  
\includegraphics[width=\textwidth]{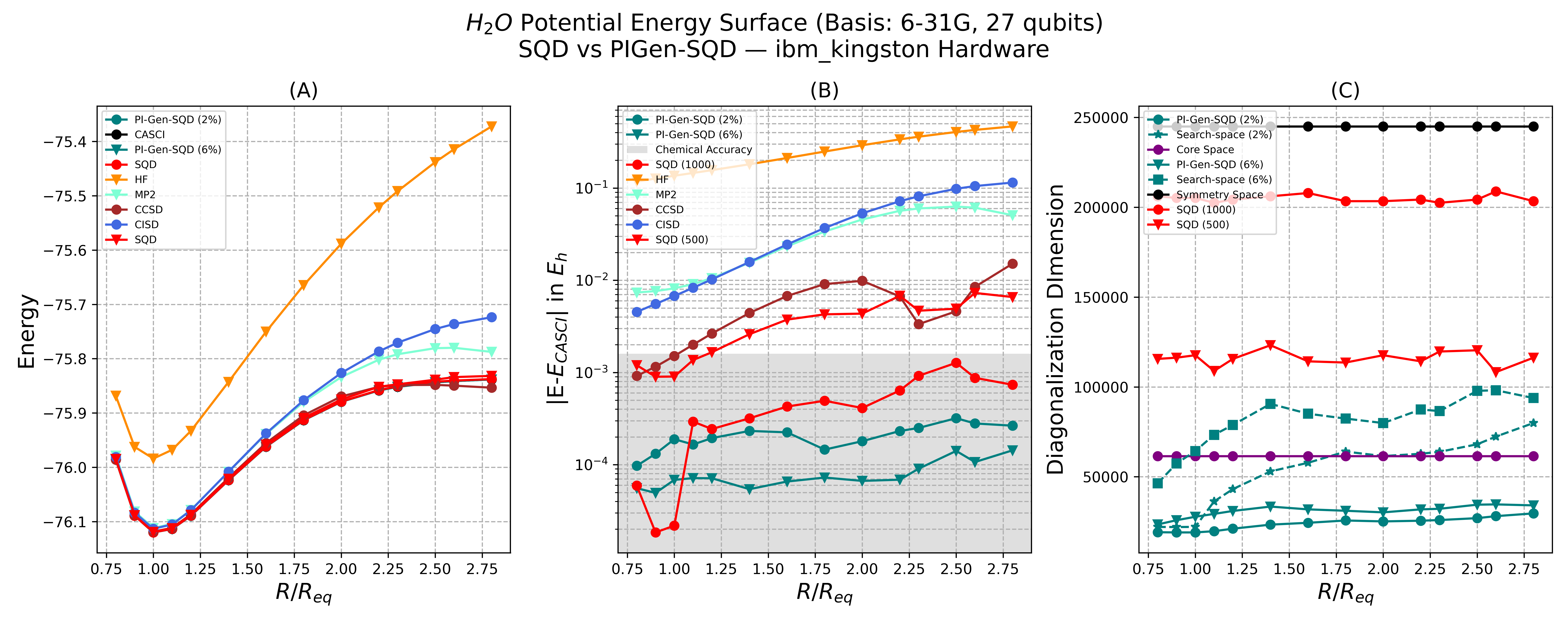}
\caption{\textbf{Potential energy curve of $\boldsymbol{\mathrm{H_2O}}$ with SQD, PIGen-SQD and other standard methods for comparison. Energy values and energy errors (in log scale) with FCI in $E_h$ are shown in y-axes panel (A) and (B) with the ratio of bond distance against equilibrium distance $\frac{R}{R_{eq}}$ plotted along x-axes of all the panels. The number of configurations in symmetry space, core space and diagonalization space in SQD and PIGen-SQD are shown in panel (C). The dotted lines in panel (C) show the search space dimension for PIGen-SQD. }}
    \label{H2O PES}
\end{figure*}

A crucial aspect of any such stochastic sampling driven subspace expansion method is the size of the \textit{search space} explored by the ML model during the self-consistent configuration reconstruction process. Owing to the inherent stochasticity in the generative step, each macro-iteration inevitably produces configurations that have negligible or no contribution to the target state. These are collected into a blacklist to prevent their re-selection in subsequent iterations. Because the target CI vector is typically sparse, this blacklist grows much faster than the dominant configuration space. As a result, the generated samples increasingly contain a significant fraction of unimportant configurations, causing the overall search space, which consists of both dominant and blacklisted configurations, to expand proportionally with iteration count.
All of these quantitative measures are shown in Fig.\ref{convergence profile plot} where evidently, the RBM based method provides almost two orders of magnitude more accurate
energy estimation while operating within a much smaller diagonalization space
compared to random sampling.
The associated search space dimension, which is a direct measure of the explored Hilbert space, is also substantially low for RBM driven generation
whereas the same for random sampling process grows much more steeply.
For random sampling to provide accurate results, it requires sufficiently large search space ($\sim d_Q$)
which is impractical from a scalability perspective.
Ideally the generative model should be trained in a way such that the blacklist space dimension is as low as possible, maximizing the
number of dominant configurations during Hilbert space exploration.
This concentrated search and configuration selection only from the relevant portion of the symmetry-restricted Hilbert space demonstrates an effective learning and
configuration generation capability of the RBM.


\subsection{Anchoring ML Model to the Target State: the Role of Physics-informed Initialization for Effective Learning}
To numerically illustrate the degree of \textit{anchoring effect}
of the physics-informed initialization provided to hardware samples, Fig. \ref{mbpt rank comparison} shows a comparative study
regarding different rank of perturbative measures having different initialization to the fermionic state reconstruction process for two different geometries
of $\mathrm{H_2O}$ and $\mathrm{N_2}$ in 6-31G basis. 
This analysis incorporates both symmetry-preserved hardware samples and perturbatively generated initializations with determinants up to ranks $n=2,3,4$. Approximate perturbative estimates at each rank were obtained by including a subset of symbolically generated algebraic terms (see Appendix \ref{PT algebraic terms}, \ref{appendix symbolic tensor contraction}).
The plot demonstrates that ML-based configuration recovery converges 
much faster with better energy accuracy when initialized with excited configurations up to rank 4
compared to only hardware, (up to) rank 2 and rank 3 initializations.
This is observed for both equilibrium and stretched geometries where strong correlation effects are more prevalent.
The inset bar plots show the number of configurations required in the \textit{search space} to 
reach chemical accuracy. In equilibrium geometries,
determinants taken up to rank 4 provide an excellent starting point and hence requires 
substantially reduced dimensional search space to reach chemical accuracy. This ensures an effective 
learning by the generative model of the distribution of the target state basis vectors.
For stretched geometries also, similar trend persists, though the relative difference between the 
search spaces narrows down compared to equilibrium geometries
(see the inset bar plots) as the presence of strong
correlation requires a substantial number of even higher rank determinants 
and more configurations in the search space. However, even in very strongly correlated regime where the low order perturbative estimates may miss out some 
important configurations, the generative learning algorithm is eventually able to pick those up, albeit with a slower macro-cyclic convergence and somewhat larger search space. Moreover, initializations with raw symmetry-preserved hardware samples show similar accuracy and
search space dimensions compared to rank 2 and rank 3 initializations in most of the cases under consideration in Fig. \ref{mbpt rank comparison}.
The benefit of augmenting hardware-derived samples with perturbatively generated configurations is illustrated in Fig.~\ref{mbpt rank comparison}(D).
In this case for the stretched geometry of $\mathrm{N_2}$, initialization with raw hardware samples (symmetry-preserved) followed by RBM-driven sampling fails to reach chemical accuracy, while PIGen-SQD initialized with various ranks of perturbatively selected configurations reaches desired accuracy consistently. 
This behavior can be attributed to the severely limited shot budget in this case ($\sim 10^6$), compared to the associated Hilbert-space dimension ($d_Q \sim 1.9 \times 10^7$), which leads to a poor-quality measurement distribution with negligible overlap with the true ground state. As a result, the RBM is unable to reliably identify the underlying structure of the wavefunction, generating only a small number of dominant determinants that are insufficient for accurate energy estimation. This example highlights the critical role of effective many-body inspired initialization in classical post-processing and demonstrates that, even under stringent shot constraints, PIGen-SQD can efficiently recover dominant configurations when appropriately initialized.
This suggests that one needs to strike the right balance between classical physics-informed initialization and the generative procedure to an accurate and efficacious energy estimation.
Toward this, from here onward, we will use $n=4$ for all PIGen-SQD calculations. However, one may choose a higher value of $n$ (with some
selected algebraic terms that scale favorably) for 
more strongly correlated systems to accelerate convergence which would be explored further in future works.

\begin{figure*}[!ht]
    \centering  
\includegraphics[width=\textwidth]{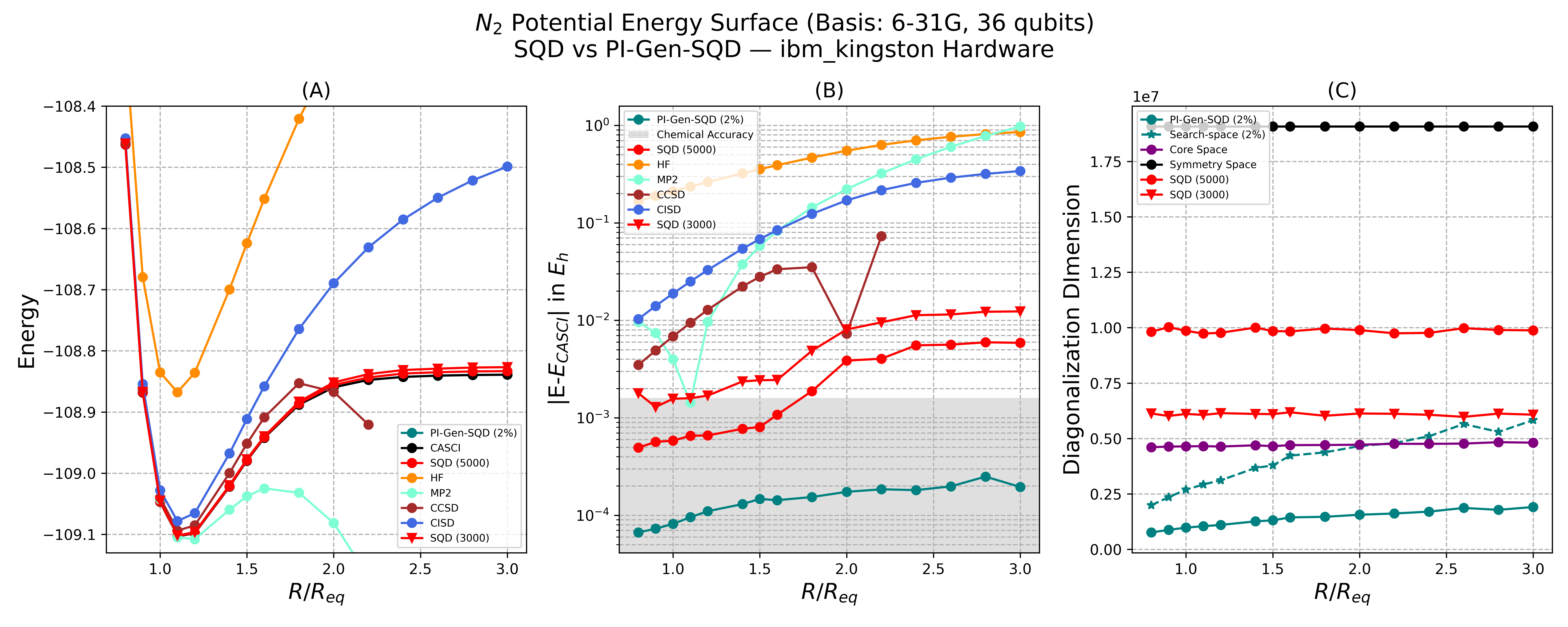}
\caption{\textbf{Potential energy curve for $\mathrm{N_2}$. The axes information are same as Fig.\ref{H2O PES}.
}}
    \label{N2 PES}
\end{figure*}

\begin{figure*}[!ht]
    \centering  
\includegraphics[width=\textwidth]{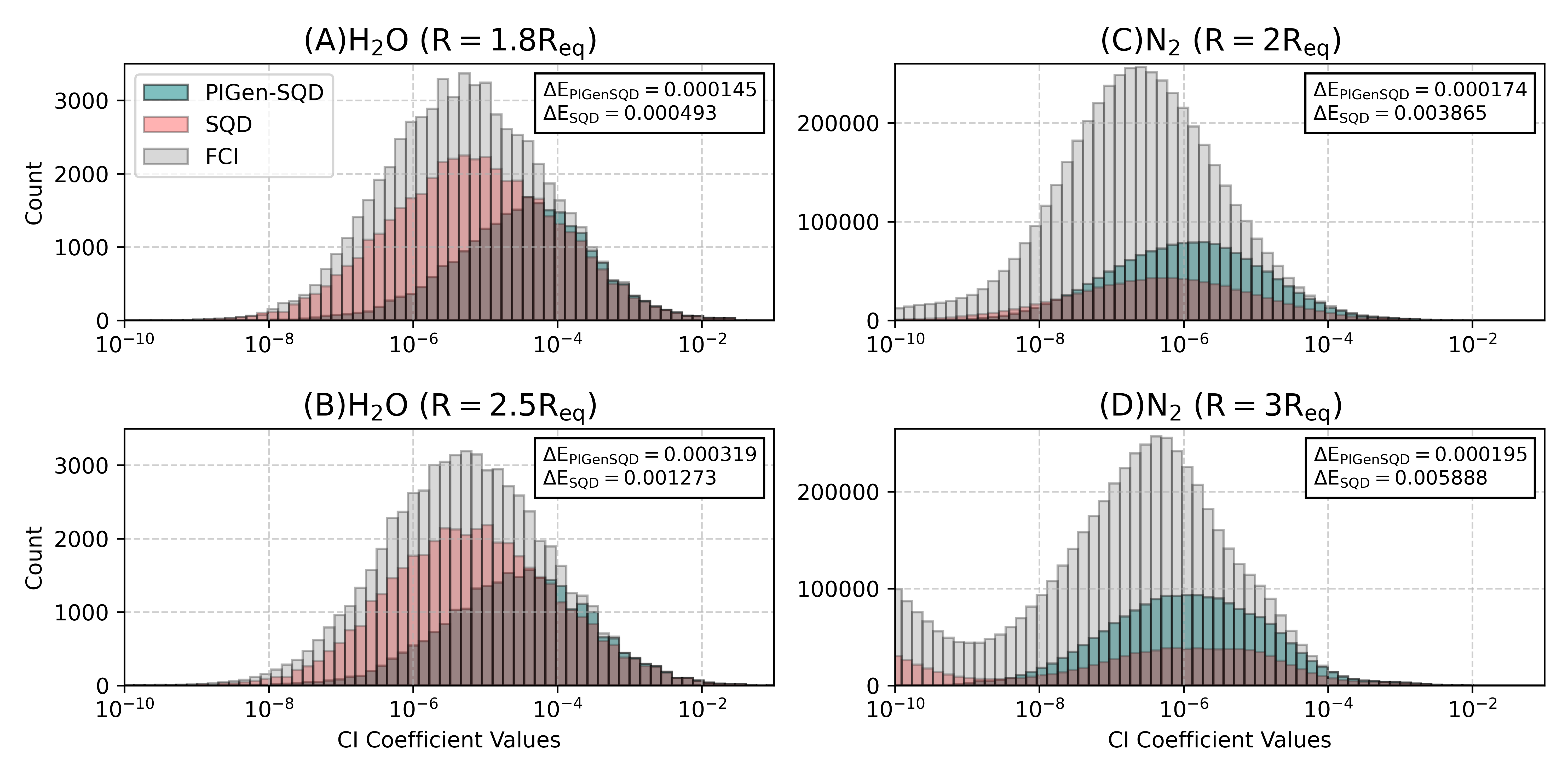}
\caption{\textbf{Distribution of the CI coefficients for PIGen-SQD, SQD and FCI. Absolute CI coefficient values from each method are sorted in ascending order and plotted along the x-axis for two geometries of $\mathbf{H_2O}$ ((A), (B)) and $\mathbf{N_2}$ ((C), (D)). For any given interval on the x-axis, the corresponding y-value indicates the number of coefficients falling within that range. PIGen-SQD more accurately reproduces the larger-magnitude CI coefficients, which provide the dominant contributions to the target wavefunction. The inset panels show the corresponding energy errors of the methods relative to FCI.
}}
    \label{ci coeff scatter plot}
\end{figure*}

\subsection{Potential Energy Profile and State-Fidelity:}
This section demonstrates the accuracy and efficiency of PIGen-SQD against conventional SQD
by comparing the performance of both the methods to predict ground state energies of $\mathrm{H_2O}$ and $\mathrm{N_2}$
in 6-31G basis
across different geometries.

$\boldsymbol{\mathrm{H_2O}:}$ The dissociation curve of $\mathrm{H_2O}$ via symmetric bond stretching in 6-31G basis set with frozen core approximation (8 electrons in 12 spatial orbitals) is shown in Fig. \ref{H2O PES} for PIGen-SQD, SQD and
other reference methods such as HF, MP2, CISD and CCSD for comparison.
Panel (A) of Fig.\ref{H2O PES} shows the potential energy surface and the associated energy difference in $E_h$ with FCI
is shown in panel (B) along y-axis with the ratio of bond distance $R$ to equilibrium distance $R_{eq}$ plotted along x-axes. The comparative study of the diagonalization dimension for SQD and PIGen-SQD is shown in panel (C)
along with the associated search space, spin-particle number conserved symmetry space and core space dimension.
SQD is performed in this case with sample size 500 and 1000 in 5 batches at each step of configuration recovery loop
while PIGen-SQD results are obtained with $x\%$ ($x=2,6$ here) of the symmetry space samples generated in each macro-cycle.
For SQD, the minimum energy and maximum diagonalization dimension among all batches are plotted.
SQD with sample size 500 or SQD(500) shows chemically accurate results near equilibrium geometry while it struggles for stretched bonds.
Since increasing sample size includes more configurations in the diagonalization subspace, SQD(1000) performs well for the entire dissociation curve. On the other hand, PIGen-SQD($2\%$) and PIGen-SQD($6\%$)
both shows results well within the chemical accuracy.
However, the main aspect that distinguishes the efficiency of SQD and PIGen-SQD is the associated diagonalization dimension.
In this study, while SQD(1000) requires almost around 84\%
of the entire symmetry space $d_Q$ for diagonalization to provide chemically accurate results for the entire potential energy surface, PIGen-SQD(2\%) does so
using only $\sim 7\%$ of the same.
Moreover, our method is systematically improvable by increasing $x$ as demonstrated in the plot with
the PIGen-SQD(6\%) that shows substantially improved accuracy compared to SQD(1000) throughout the dissociation curve particularly in the strongly correlated region.
Another important metric of our method is the search space that
RBM explores during the self-consistent configuration recovery process. From Fig.\ref{H2O PES} it is evident that
the search space for PIGen-SQD is around the core space dimension for $x=2\%$ for the entire dissociation curve while it slightly increases for $x=6\%$ providing better accuracy.
However, in all the cases, even the search space dimension is well below that of SQD diagonalization dimension, showing
an effective prediction of the underlying distribution of the dominant states in the Hilbert space.
While PIGen-SQD can be systematically improved by tuning $x$, in our subsequent studies, we will use $x=2\%$ as it strikes an optimal balance between the accuracy and the explored Hilbert space.

\begin{figure*}[!ht]
    \centering  
\includegraphics[width=\textwidth]{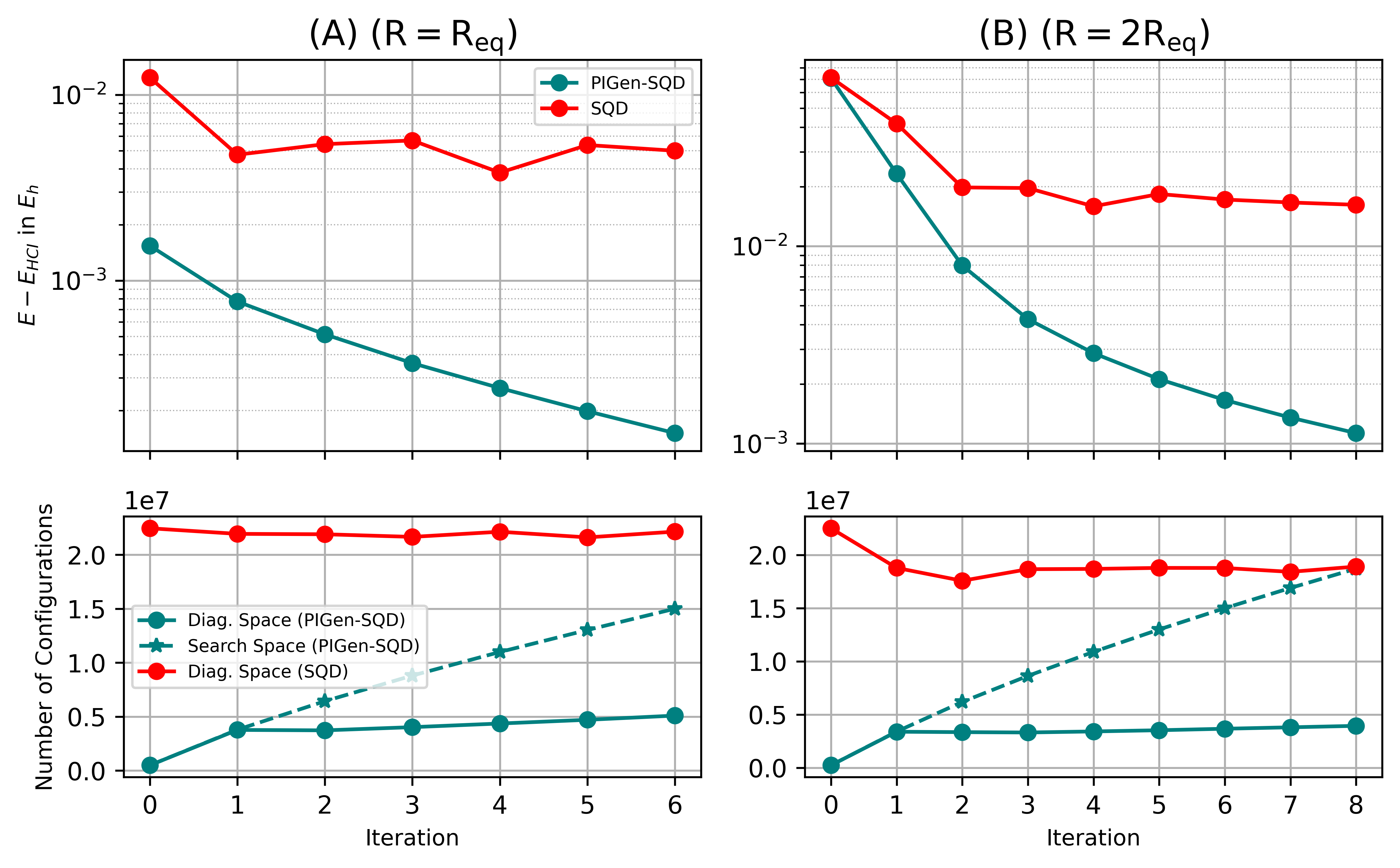}
\caption{\textbf{SQD v/s PIGen-SQD iterations for two geometries of $\boldsymbol{\mathrm{C_2H_2}}$ (45 qubits system). In the first row energy errors with respect to HCI is shown while
the second row shows the associated diagonalization subspace and the search space dimension. Evidently PIGen-SQD shows more than one order-of-magnitude more optimal
energy estimation compared to SQD with with 75$\%$ reduction in the associated diagonalization subspace dimension.
}}
    \label{C2H2 convergence}
\end{figure*}

$\boldsymbol{\mathrm{N_2}:}$ Bond stretching of the $\mathrm{N_2}$ ($R_{eq}=1\mbox{\AA}$) molecule is a paradigmatic approach to demonstrate electronic strong correlation effects.
Fig. \ref{N2 PES} shows the dissociation curve of $\mathrm{N_2}$ for both SQD and PIGen-SQD along with other standard methods in 6-31G basis with frozen core approximation resulting in 10 electrons in 16 spatial orbitals. The SQD results are obtained using sample size 3000 and 5000 with fixed batch size 5. For some stretched geometries CCSD did not converge and those points are not shown in the figure. While both SQD(3000) and SQD(5000) shows
chemically accurate results near equilibrium geometries, it fails in stretched geometries having errors of the order of $10mE_h$ and $5mE_h$
with around 6 and 10 million determinants in the diagonalization space respectively. Contrarily, PIGen-SQD(2\%) shows drastically improved results maintaining
around $0.015mE_h$ accuracy throughout the dissociation curve, even under strong correlation at stretched geometries.
Remarkably, PIGen-SQD needs only about 1-1.5 million determinants in the diagonalization subspace, which shows almost an order-of-magnitude
reduction in the diagonalization dimension while providing more than an order-of-magnitude energy accuracy.
The associated search space is well under the core space dimension near equilibirum geometries,
whereas for stretched geometries it grows to be comparable in size to the core-space dimension.
However, the search space is still well below the diagonalization dimension of SQD(5000).

    

\textbf{State Fidelity:}

Fig. \ref{ci coeff scatter plot} illustrates an approximate estimation of fidelity of the reconstructed wavefunctions by comparing the CI coefficients obtained from the different methods against FCI coefficients.
The absolute CI coefficients obtained from FCI, SQD and PIGen-SQD are sorted in an ascending order and plotted along the x-axis for the two geometries of $\mathrm{H_2O}$ ((A), (B)) and $\mathrm{N_2}$ ((C), (D)). For any given interval on the x-axis, the corresponding y-value indicates the number of coefficients falling within that range. The corresponding energy difference from FCI is given in the inset text for all the cases.
For $\mathrm{H_2O}$, both SQD and PIGen-SQD shows similar order of accuracy though the number of configurations
predicted by SQD is substantially large. Fig.\ref{ci coeff scatter plot} (A, B) histograms show PIGen-SQD distribution has a peak toward the larger CI
coefficients ($\sim\mathrm{10^{-4}}$) representing a dominant subset of the FCI coefficients that provides major support to target state and contribute more toward accurate
energy estimation. Better state fidelity for PIGen-SQD is more conspicuous for stretched geometries of $\mathrm{N_2}$ in Fig. \ref{ci coeff scatter plot}
(C, D) as one can see the substantially higher peak of PIGen-SQD curve toward the region with larger CI coefficients than SQD which has a flatter distribution over the entire range of values. This is reflected in the corresponding energy errors (see the inset text) where PIGen-SQD shows more than one order of magnitude better energy accuracy ($\sim$0.1$mE_h$ compared to $\sim$4$mE_h$ obtained by SQD) under strong correlation.
These examples demonstrate that PIGen-SQD faithfully reconstructs accurate high-fidelity states as compared to SQD
leading to significantly improved energy predictions with substantially reduced requirements of HPC resources.
The corresponding density plot in Fig.\ref{density plot} (see Appendix \ref{appendix density plot}) shows the predicted coefficients are sharply peaked around
larger values with maximum probability for PIGen-SQD, making it a unimodal distribution. On the contrary, SQD and FCI curves are
bimodal with a higher peak near practically zero coefficients due to sparsity.
This further confirms the concentrated sampling of the dominant sector of the Hilbert space enabled by PIGen-SQD.

\subsection{Beyond Exact Diagonalization}


\begin{figure*}[!ht]
    \centering  
\includegraphics[width=\textwidth]{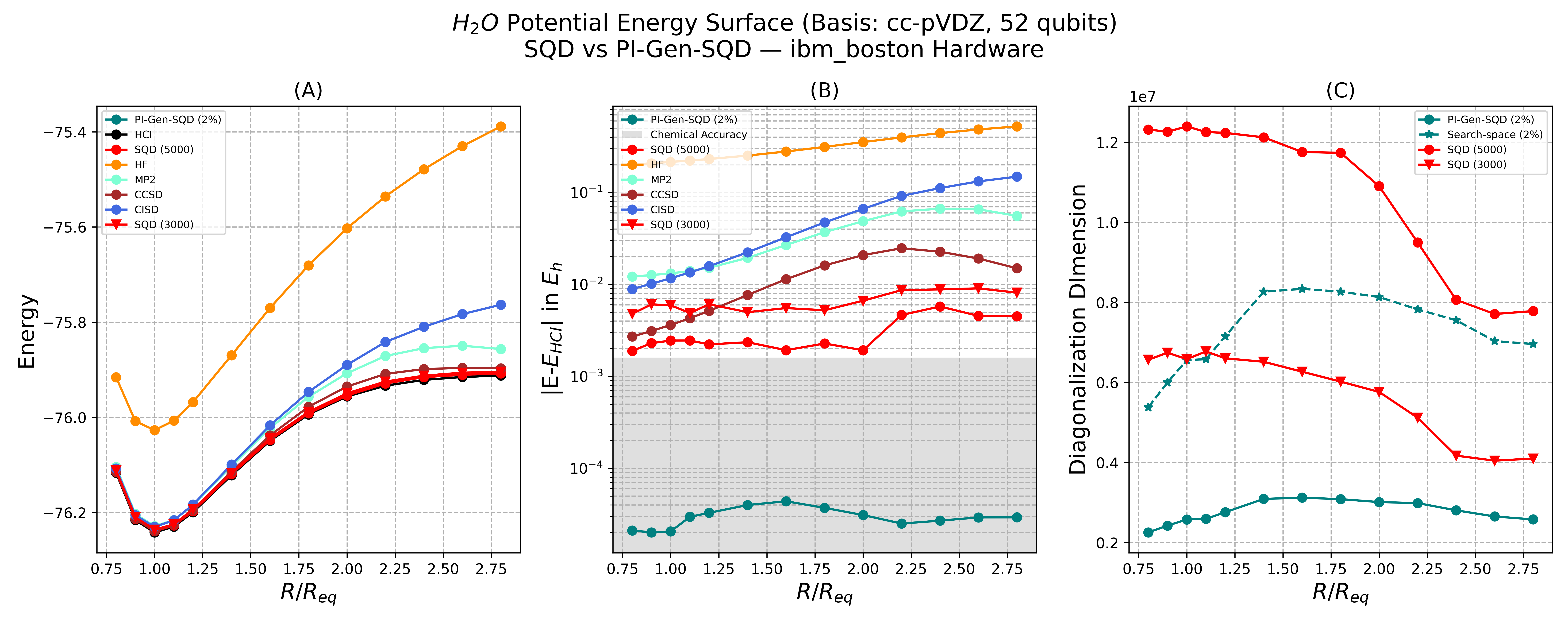}
\caption{\textbf{Dissociation curve for $\mathrm{H_2O}$ in cc-pVDZ basis. The axes information is same as Fig. \ref{H2O PES}}}
    \label{PES plot H2O ccpvdz}
\end{figure*}

Since the Hilbert space grows exponentially with system size, exact diagonalization becomes infeasible for larger molecular systems and extended basis sets. We therefore consider two representative cases $\mathrm{C_2H_2}$ in the 6-31G basis and $\mathrm{H_2O}$ in the cc-pVDZ basis for which exact diagonalization is computationally impractical. In these regimes, we assess the performance of PIGen-SQD in comparison with standard SQD.

Fig. \ref{C2H2 convergence} shows PIGen-SQD versus SQD convergence for two isolated single point calculations of $\mathrm{C_2H_2}$ (C-C equilibrium bond distance $R_{eq} = 1.202\mbox{\AA}$).
SQD configuration recovery calculations are performed in this case using sample size 5000 in 5 batches.
Each column in Fig.\ref{C2H2 convergence} shows the energy difference with HCI in log-scale (top row) and the number of configurations required (bottom row)
in diagonalization subspace and search space of the respective methods against a common x-axis
representing number of iterations. Column (A) and (B) shows calculations for C-C bond-length near equilibrium geometry $R=R_{eq}$ and stretched geometry $R=2R_{eq}$.
PIGen-SQD convergence behaviour shows iteratively improved energy with more than one-order-of-magnitude
better accuracy compared to SQD while keeping the diagonalization dimension and search space as low as possible.
Quantitatively, even with a 75$\%$ reduction in the associated diagonalization subspace dimension for both the cases under consideration,
energy accuracy of around 0.1$mE_h$ and 1$mE_h$ is achieved for equilibrium and stretched geometries compared to 5$mE_h$ and 10$mE_h$
for SQD, showing the superiority of PIGen-SQD under strong correlations.

Dissociation curve of $\mathrm{H_2O}$ with cc-pVDZ basis (with the frozen core approximation) is shown in Fig. \ref{PES plot H2O ccpvdz}.
This is a challenging case with the associated symmetry space dimension $d_Q\sim 7.8\times10^7$.
With sub-optimal $10^6$ shots, SQD struggles to identify the dominant configurations
in this case as well and shows $\sim2mE_h$ energy accuracy with SQD(5000) for all the geometries studied here compared to HCI.
Contrarily PIGen-SQD provides consistently
more accurate energy estimations ($\sim10^{-2}mE_h$) with only around 10$\%$ of the diagonalization cost of SQD.

It is important to mention in this aspect that while in all the cases under consideration PIGen-SQD
shows up to 90$\%$ reduction in the diagonalization subspace dimension along with orders-of-magnitude more accurate energy estimations,
the performance could be systematically improved further with increased shot budget, optimized hyper parameters and more robust
machine learning models such as transformers.\cite{shang2025solving}

\section{Conclusion and Future Outlook}
In this paper, we propose PIGen-SQD, a novel QCSC workflow based on SQD framework that leverages a strategic orchestration of
sampling efficiency of quantum computers,
many-body intuitions and generative power of machine learning to provide better convergence with substantially reduced classical resource utilization.
QCSC/ SQD utilizes the quantum computer as a sampling engine to generate configuration subspace in which the molecular Hamiltonian is projected and diagonalized.
In the presence of quantum hardware noise, SQD requires a configuration recovery module, a heuristic technique that \enquote{corrects} the noise-affected samples with iterative
filtering based upon the correct spin-particle number symmetry and average occupation probability. Accuracy of SQD relies upon the
the degree of initial overlap with the target state provided by the sampled distribution and the capability of heuristic configuration recovery
scheme to predict dominant configurations. However, the key challenge is that the state preparation on current quantum hardware does not necessarily
guarantee high quality samples to provide this support to the target state and may require prohibitively huge number of shots which collectively
adds a heuristic component to the SQD pipeline. To address this,
here we introduce a physics-informed pre-processing which employs perturbative measures with efficiently designed implicit low rank tensor decompositions to identify a particular group of dominant
configurations. These perturbative measure when combined with hardware samples, yields a substantially improved initial overlap
to the target state providing a head-start for configuration recovery. This is followed by integration of generative ML models
(RBM in this case) which upon training with this sample distribution, generates new configurations while exploring only
the relevant sector of the Hilbert space of chemical relevance.
Our numerical demonstrations with 27, 36, 45 and 52 qubit molecular simulations
conducted on state-of-the-art IBM Heron R2 and R3 processors along with HPC resources
suggest PIGen-SQD outperforms traditional SQD in both accuracy and efficiency, providing up to one-order of better energy
accuracy with up to 90$\%$ reduction in the corresponding subspace dimension required for Hamiltonian diagonalization.
Such accuracy with substantially reduced HPC resource requirements is observed even under strong electronic correlations where SQD
struggles to predict chemically accurate results.

From a near-future application perspective, PIGen-SQD can be integrated with other SQD variants such as Sample-based Krylov Quantum Diagonalization (SKQD)\cite{yoshioka2025krylov}
or its more efficient variant using quantum stochastic drift protocol (SqDRIFT)\cite{piccinelli2025quantum} for guaranteed and faster convergence along with enhanced
computational efficiency.
Moreover, the current architecture have huge scope of technical improvements such as RBM can be replaced with more advanced neural network models such as transformers\cite{shang2025solving} which has recently been shown to provide unique advantages while dealing with complex molecular systems.
With continued improvements in quantum-hardware capabilities and machine-learning components, PIGen-SQD presents a promising pathway toward scalable molecular simulations, advancing the field toward resolving several longstanding challenges in chemistry.







\section*{Author contributions}

\textbf{Conceptualization}: Chayan Patra, Rahul Maitra

\textbf{Methodology}: Chayan Patra, Rahul Maitra

\textbf{Investigation:} Chayan Patra, Rahul Maitra,  Dibyendu Mondal

\textbf{Formal analysis:} Chayan Patra, Rahul Maitra, Dibyendu Mondal

\textbf{Data curation:} Chayan Patra

\textbf{Writing - original draft:} Chayan Patra, Rahul Maitra

\textbf{Writing - review and editing:} All authors

\textbf{Software:} Chayan Patra, Mostafizur Rahaman Laskar, Sonaldeep Halder, Richa Goel

\textbf{Visualization:} Chayan Patra

\textbf{Supervision:} Rahul Maitra

\textbf{Funding acquisition:} Rahul Maitra

\section*{Conflicts of interest}
The authors have no conflict of interest to disclose.

\section*{Data and Code Availability}
The data and code are available upon reasonable request to the corresponding author.

\section*{Acknowledgments}
We acknowledge the use of IBM Quantum Credits for this work. 
The authors acknowledge Ritajit Majumdar from the Quantum Algorithm Engineering team at IBM for fruitful discussions, and contributions to achieving results on quantum hardware. RM acknowledges the financial support from Industrial Research 
and Consultancy Centre (IRCC), IIT Bombay and Anusandhan National Research Foundation (ANRF, erstwhile SERB), Government
of India (Grant Number: MTR/2023/001306).
CP acknowledges University Grants Commission (UGC) for the fellowship.
DM and SH thank Prime Minister's Research Fellowship (PMRF) and Council of Scientific and Industrial Research (CSIR) for their
respective fellowships.






\appendix

\section*{Appendix}

\begin{figure*}[!ht]
    \centering  
\includegraphics[width=\textwidth]{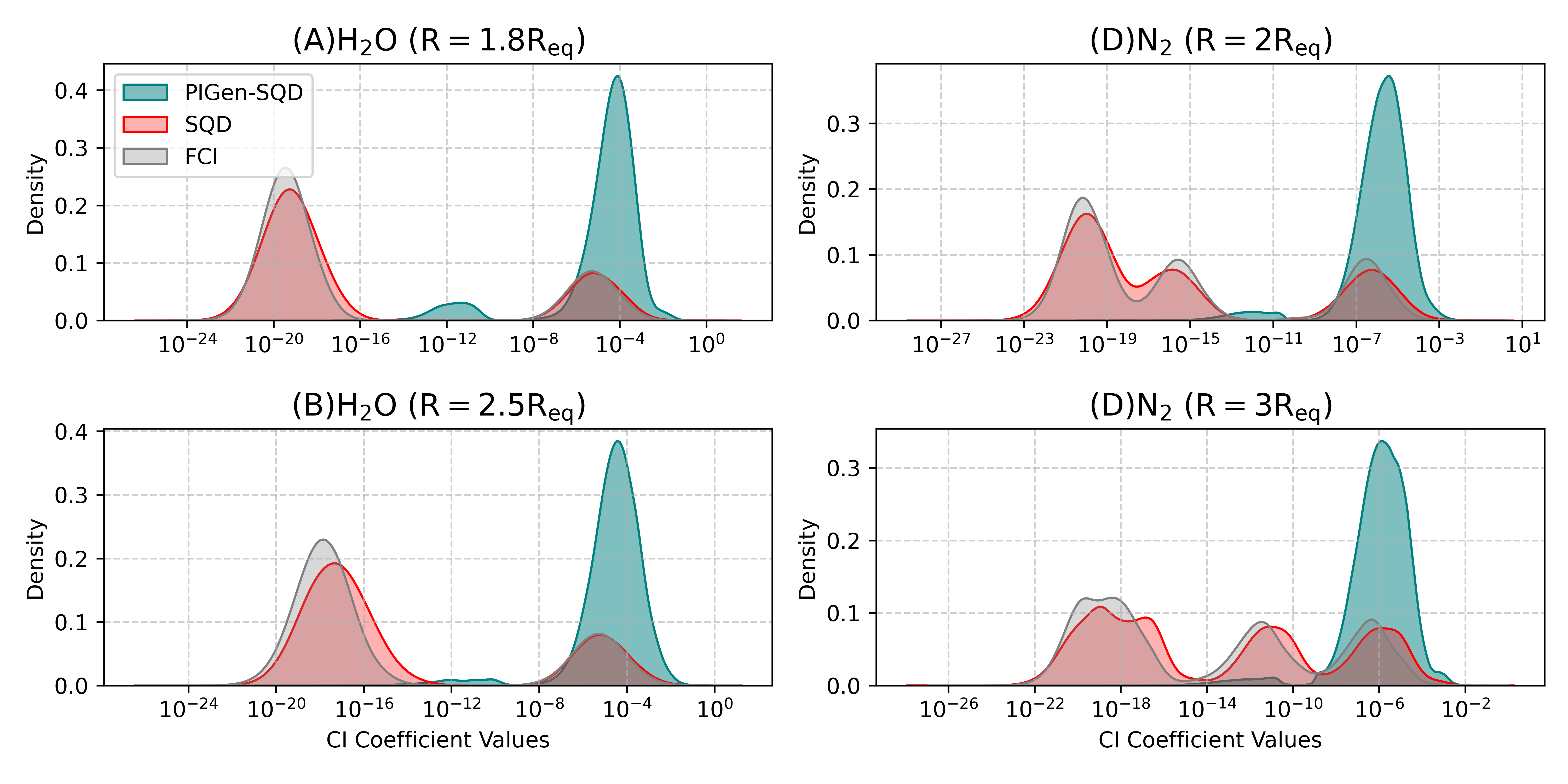}
\caption{\textbf{Normalised density of the CI coefficients for PIGen-SQD, SQD and FCI. Absolute CI coefficient values from each method are sorted in ascending order and plotted along the x-axis for two geometries of $\mathbf{H_2O}$ ((A), (B)) and $\mathbf{N_2}$ ((C), (D)).
Here y-axes show the corresponding kernel density estimate which converts the data shown in Fig. \ref{ci coeff scatter plot}
into a probability density.
The unimodal behaviour of PIGen-SQD over bimodal SQD shows that PIGen-SQD performs concentrated search while SQD predicted
configurations contain many unimportant configurations.
}}
    \label{density plot}
\end{figure*}

\section{Algebraic Terms for Perturbative Calculations} \label{PT algebraic terms}
The perturbative calculations with $n=3$ and $n=4$ configurations include the following algebraic terms:

\begin{equation} \label{tensor contraction terms for triples}
\begin{split}
    & c_{ijk}^{abc} \leftarrow \{v_{ij}^{am}t_{mk}^{bc}, v_{ie}^{ab}t_{jk}^{ec}\}
\end{split}
\end{equation}

\begin{equation} \label{eq: c4 terms}
\begin{split}
    c_{ijkl}^{abcd} \leftarrow & \{v_{ij}^{am} v_{mk}^{bn} t_{nl}^{cd},  v_{ie}^{ab} v_{jf}^{ec} t_{kl}^{fd}, v_{ij}^{am} v_{mf}^{bc} t_{kl}^{fd},  v_{ie}^{ab} v_{jk}^{en} t_{nl}^{cd},\\
    &  v_{ij}^{am} t_{mn}^{bc} v_{kl}^{nd},  v_{ie}^{ab} t_{jk}^{ef} v_{fl}^{cd},  v_{ij}^{am} t_{mk}^{bf} v_{fl}^{cd}\}
\end{split}
\end{equation}

Note the signs of the algebraic terms are not explicitly shown in these expressions are are assumed to be absorbed within each term.

The triply excited configurations arising from second-order perturbative corrections to the wave function are generated by the two terms in Eq. \eqref{tensor contraction terms for triples}. Among these, the dominant computational bottleneck is the term $v_{ie}^{ab}t_{jk}^{ec}$, which involves contraction over a virtual orbital index. In practice, a given triply excited configuration can be generated through multiple contraction pathways. Consequently, the excitation manifolds produced by the $v_{ij}^{am}t_{mk}^{bc}$
and $v_{ie}^{ab}t_{jk}^{ec}$ terms exhibit substantial overlap. Since the objective of the PIGen-SQD workflow is to provide a support to quantum-hardware-derived samples rather than to construct a complete perturbative expansion, the perturbative evaluation can be selectively restricted to those terms with more favorable computational scaling based on available classical HPC resources.

With the increase in the order of the perturbative correction,
the number of terms contributing to a particular excitation manifold grows rapidly as shown in Eq. \eqref{eq: c4 terms} for quadruples generations.
In such cases also, the computations can be restricted to only a few terms if necessary as the PIGen-SQD workflow does not
demand an exact calculation of all the terms in the expansion.

\section{Illustration of Symbolic Tensor Operation} \label{appendix symbolic tensor contraction}

Given the terms $v_{i_se_s}^{a_sb_s}$ and $t_{j_tk_t}^{e_tc_t}$, which belong to $\hat{S}_I$ and $\hat{T}_{\mu_2}$ respectively and are greater than the threshold $\epsilon_{int}$, the symbolic tensor manipulations for triples generation can be carried out as follows:

\begin{itemize}
    \item Group the unique outer non-contractible indices $((i_s a_s b_s), (\{e_s\}))$ and $((j_t k_t c_t), (\{e_t\}))$, while allowing the contractible indices ($e_*$) to vary within each group.
    
    \item For a pair of $(i_s a_s b_s)$ and $(j_t k_t c_t)$, if exactly one possible value in $\{e_s\}$ matches one in $\{e_t\}$, retain the corresponding outer indices pair to form a triple $(i_s j_t k_t a_s b_s c_t)$, subject to the conditions $k_t > j_t > i_s$ and $c_t > b_s > a_s$ in order to avoid double counting.
    
\end{itemize}

This leads to a partial tensor contraction and one does not have to verify or sum over all the contractible indices, as their precise numerical values are not particularly crucial in this context. Since both $v_{**}^{**}$ and $t_{**}^{**}$ are greater than a threshold, the resultant triple would likely to be significant. While this procedure is not numerically exact, it eliminates the need for an explicit summation over the contractible indices and thereby reduces the computational effort in practice.


\section{Density Plot of the Predicted CI Coefficients} \label{appendix density plot}

A kernel density estimate (KDE) plot of the underlying probability distribution associated with the predicted CI coefficients are shown in Fig.\ref{density plot}. This density representation clearly reveals the distinct statistical behavior of the predicted CI coefficients across methods.
For PIGen-SQD, the density curve exhibits a single, sharply defined peak concentrated at relatively large coefficient values, reflecting a unimodal distribution. This indicates that the algorithm consistently identifies and samples configurations with substantial weight in the true wave function.
In contrast, both SQD and FCI display bimodal density profiles, characterized by two well-separated peaks: a prominent one clustered near negligible (practically zero) coefficients stemming from the inherent sparsity of the configuration space and a secondary peak corresponding to the few dominant configurations.
Such contrasting modal structures highlight the selective sampling behavior of PIGen-SQD which produces a markedly more focused and thus more informative representation of the dominant determinants.


\begin{figure*}
    \centering  
\includegraphics[width=\textwidth, height=10cm]{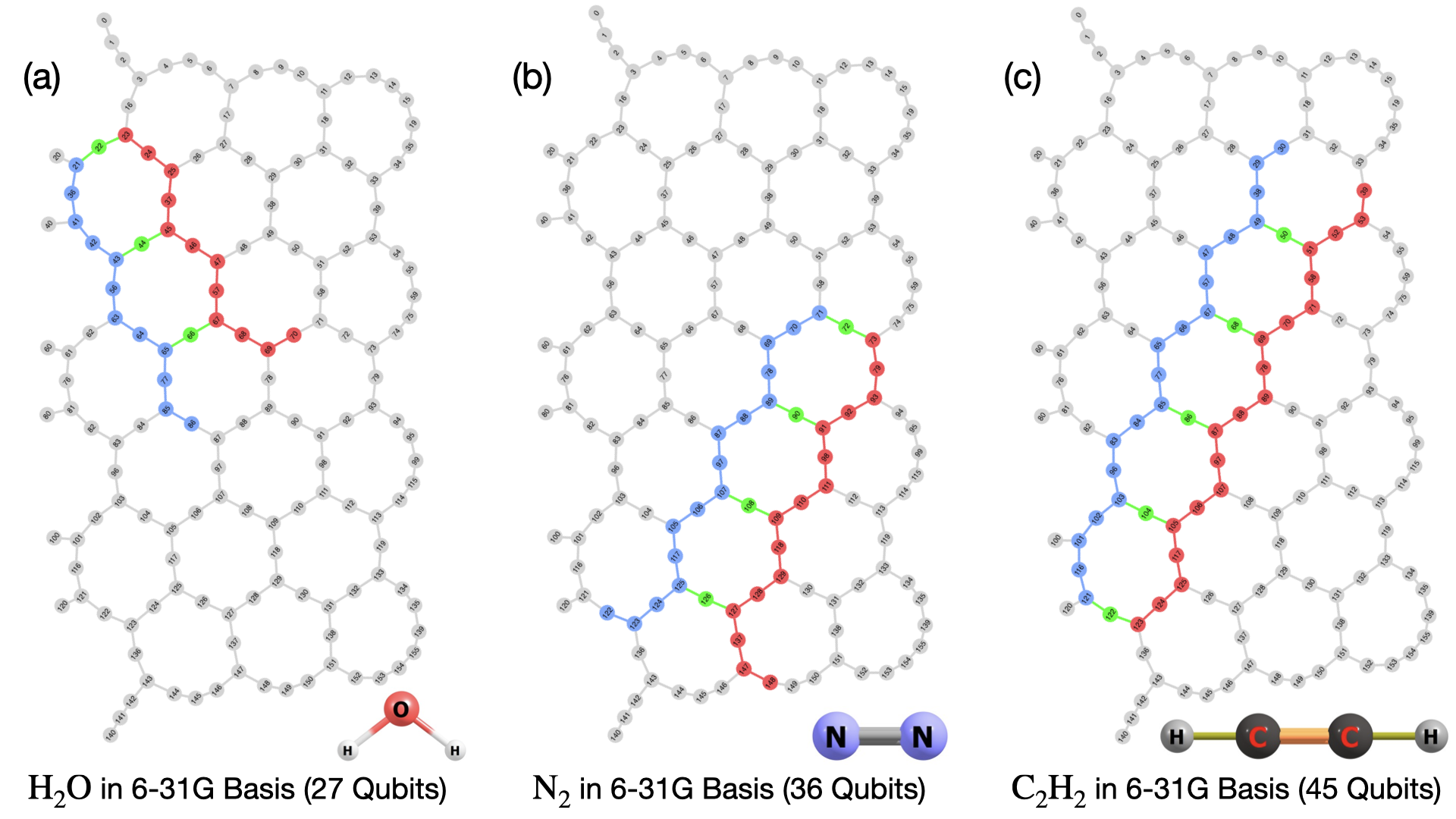}
\caption{\textbf{ Hardware coupling map layout and qubit assignments for $\boldsymbol{\mathrm{H_2O}}$ (27 qubits), $\boldsymbol{\mathrm{N_2}}$ (36 qubits) and $\boldsymbol{\mathrm{C_2H_2}}$ (45 qubits) on
\textit{ibm\_kingston} (156 qubits IBM Heron R2 processor).
Each circle represents a qubit and is assigned a number written at the center of it.
Alpha and beta molecular orbitals are mapped onto red and blue colored qubits respectively. The green colored qubits are ancilla qubits that connect adjacent alpha and beta
qubits.} 
}
    \label{hardware layout}
\end{figure*}

\section{Hardware Layout and Qubit Assignments for the Experiments Performed on IBM Heron R2 and R3 Processors} \label{appendix: hardware layout and coupling map}

\begin{figure*}[!ht]
    \centering  
\includegraphics[width=\linewidth]{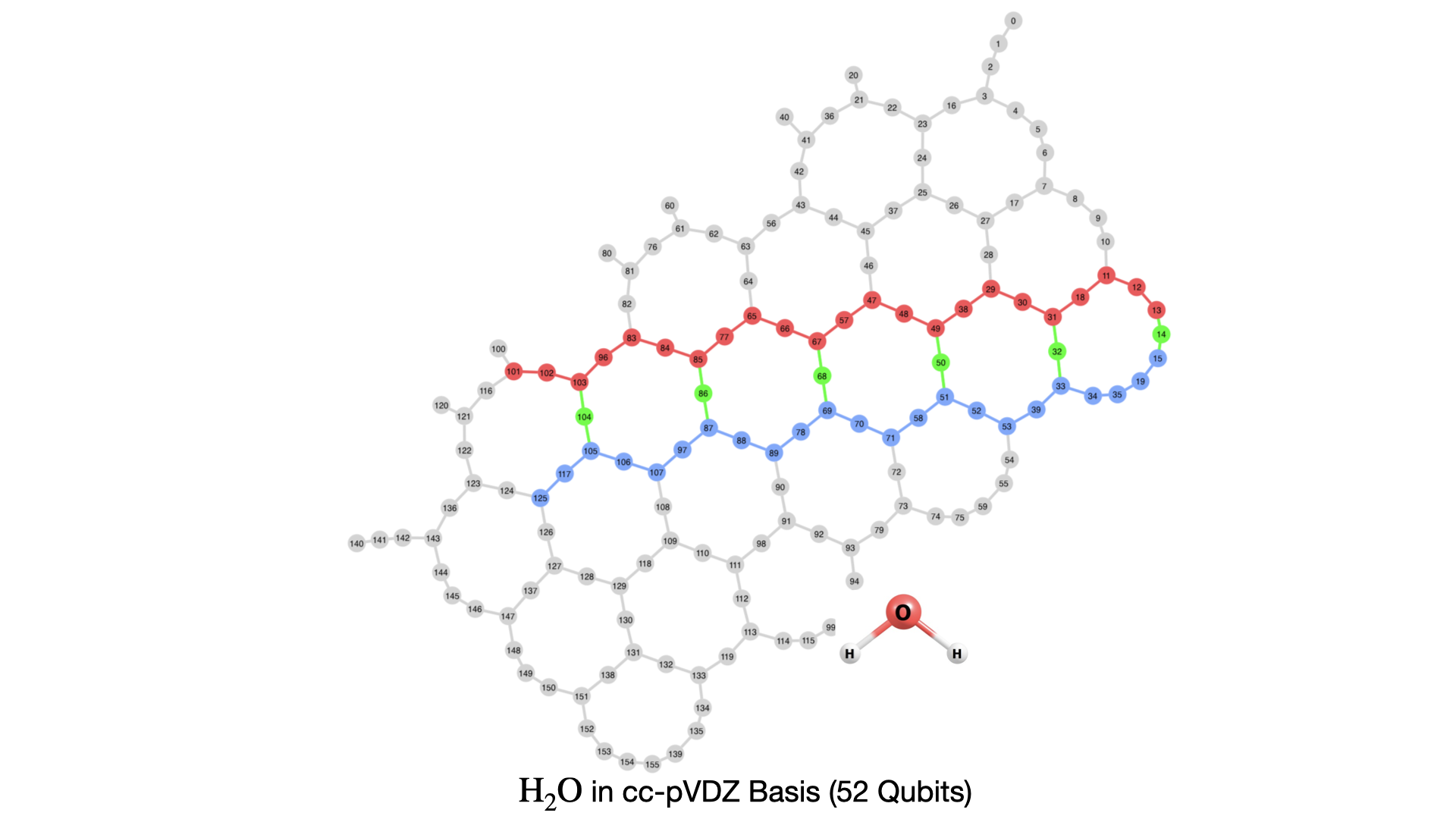}
\caption{\textbf{ Hardware coupling map layout and qubit assignments for $\boldsymbol{\mathrm{H_2O}}$ in cc-pVDZ basis (52 qubits) on
\textit{ibm\_boston} (156 qubits IBM Heron R3 processor).} 
}
    \label{hardware layout R3}
\end{figure*}

The numerical experiments on $\mathrm{H_2O}$, $\mathrm{N_2}$ and $\mathrm{C_2H_2}$ are performed on \textit{ibm\_kingston} device (156 qubits IBM Heron R2 processor).
The corresponding heavy-hex coupling map and qubit assignments for each experiments are shown in Fig. \ref{hardware layout} for 27, 36 and 45 qubits respectively.
Each small circle represent a qubit with red and blue coloured qubits representing mapped alpha and beta spin orbitals respectively.
Green coloured qubits are auxiliary qubits that connect adjacent alpha and beta qubits as per the structure of the LUCJ ansatz.
The coupling map for IBM Heron R3 processor for 52 qubits $\mathrm{H_2O}$ (cc-pVDZ basis) experiment is shown in Fig. \ref{hardware layout R3}.

\clearpage


\section*{References}


\providecommand{\noopsort}[1]{}\providecommand{\singleletter}[1]{#1}%
%


\end{document}